\documentclass[fleqn,usenatbib]{mnras}

\usepackage{newtxtext,newtxmath}

\usepackage[T1]{fontenc}

\DeclareRobustCommand{\VAN}[3]{#2}
\let\VANthebibliography\thebibliography
\def\thebibliography{\DeclareRobustCommand{\VAN}[3]{##3}\VANthebibliography}

\usepackage{graphicx}	
\usepackage{amsmath}	
\usepackage[dvipsnames]{xcolor}

\newcommand{\MSUN}{{\rm M}_{\sun}}
\newcommand{\MHOST}{M_{\rm 200c,\,host}}
\newcommand{\RHOST}{r_{\rm 200c,\,host}}
\newcommand{\RSB}{r_{\rm *,SB=26.5}}
\newcommand{\RHALF}{r_{\rm *,1/2}}
\newcommand{\RHALFV}{r_{\rm V,1/2}}
\newcommand{\RHALFGEN}[1]{r_{\rm *,1/2,#1}}
\newcommand{\MSTAR}{M_{\rm *}}

\newcommand{\XXC}{\textbf{SM2}}
\newcommand{\XVC}{\textbf{SM1}}
\newcommand{\XC}{\textbf{FM}}
\newcommand{\XCX}{\textbf{LM1}}
\newcommand{\XCXX}{\textbf{LM2}}
\newcommand{\XCRII}{\textbf{FM-R2}}
\newcommand{\XCRIII}{\textbf{FM-R3}}
\newcommand{\targetMerger}{$z \approx 2$}
\newcommand{\vgn}{VG}
\newcommand{\tng}{TNG}

\title[How merger histories shape galaxy sizes]{The PARADIGM project I: How early merger histories shape the present-day sizes of Milky-Way-mass galaxies}

\author[G. D. Joshi et al.]{Gandhali D. Joshi,$^{1,2}$\thanks{E-mail: gandhali.d.joshi@durham.ac.uk}
Andrew Pontzen,$^{1,2}$
Oscar Agertz,$^{3}$
Martin P. Rey,$^{4,5}$
Justin Read$^{6}$ \&
Annalisa Pillepich$^{7}$
\\
$^{1}$Department of Physics and Astronomy, University College London, Gower St., London WC1E 6BT, UK \\
$^{2}$Institute for Computational Cosmology, Durham University, Lower Mountjoy, South Rd, Durham DH1 3LE, UK \\
$^{3}$Lund Observatory, Division of Astrophysics, Department of Physics, Lund University, Box 43, SE-221 00 Lund, Sweden \\
$^{4}$Sub-department of Astrophysics, University of Oxford, DWB, Keble Road, Oxford OX1 3RH, UK \\
$^{5}$Department of Physics, University of Bath, Claverton Down, Bath BA2 7AY, UK \\
$^{6}$Department of Physics, University of Surrey, Guildford GU2 7XH, UK \\
$^{7}$Max Planck Institute f\"{u}r Astronomie, Königstuhl 17, D-69117 Heidelberg, Germany \\
}

\date{Accepted XXX. Received YYY; in original form ZZZ}

\pubyear{2025}

\begin{document}
\label{firstpage}
\pagerange{\pageref{firstpage}--\pageref{lastpage}}
\maketitle

\begin{abstract}
How mergers affect galaxy formation depends on both feedback processes, and on the geometry and strength of the mergers themselves. We introduce the PARADIGM project, where we study the response of a simulated Milky-Way-mass galaxy ($M_{\rm 200c}\sim 10^{12}\MSUN$ at $z=0$) forming in a cosmological setting to differing merger histories, using genetically modified initial conditions, each simulated with the VINTERGATAN and IllustrisTNG codes. While VINTERGATAN has been developed with an emphasis on resolving the cold interstellar medium, IllustrisTNG uses a subgrid two-phase model and consequently scales to large volume simulations, making them ideal to examine complementary views on how merger histories and feedback interact. Our genetic modifications alter the mass ratio of an important \targetMerger{} merger while maintaining the halo's $z=0$ mass. Whether simulated with VINTERGATAN or IllustrisTNG, smaller mass ratios for this early merger result in larger galaxies at $z=0$, due to a greater build up of a kinematically cold disc. We conclude that such broad trends are robustly reproducible; however, the normalization of the resulting stellar sizes is substantially different in the two codes (ranging between $0.5-1.7\ \rm{kpc}$ for VINTERGATAN but $1.3-7.0\ \rm{kpc}$ for IllustrisTNG). The VINTERGATAN galaxies systematically form stars earlier, leading to a larger bulge component. Despite the difference in size normalization, both simulation suites lie on the observed size-mass relation for their respective morphological types. In light of these results, we discuss the interplay between internal processes and large scale gravitational interactions and gas accretion, and how the two galaxy models converge on similar emergent trends but along different evolutionary pathways.
\end{abstract}




\section{Introduction}

Mergers of dark matter (DM) haloes and the galaxies embedded in them play a crucial role in galaxy evolution. For example, galaxy mergers may trigger starbursts in the merging systems \citep[BHs; e.g.][]{Kaufmann2000,Springel2005,Hopkins2006,Lin2007,Schawinski2010,Snyder2011,Knapen2015,Moreno2015,Renaud2022} while simultaneously allowing more efficient transport of gas towards the centres of merging galaxies, thereby boosting the growth of their central black holes \citep[BHs; e.g.][]{DiMatteo2005,Springel2005,Hopkins2006,Schawinski2010,Pontzen2017,Davis2019}. Mergers may also lead to the quenching of galaxies after such a starburst event due to the rapid consumption of gas \citep[e.g.][]{Peng2010,Gabor2010,Puglisi2021,Davies2022} or due to morphological transformation \citep{Peterson2023}. Mergers are also expected to be the most important mechanism for transforming discy, rotationally-supported galaxies into elliptical, dispersion-dominated galaxies \citep[e.g.][]{Barnes1996,Cox2006,Kormendy2009}. Beyond the above processes however, there are several open questions about the impact of mergers on galaxies, especially on longer, multi-Gyr timescales. Understanding how the overall merger history of a galaxy shapes its present day properties is key to a comprehensive understanding of galaxy evolution.

One way in which galaxies may retain memory of their merger history is through their radial extent. Observed and simulated galaxies span a broad distribution in the size-mass plane \citep[e.g.][]{Shen2003,Bernardi2010,Lange2015,TNG50Pillepich2019,Du2024} as a result of diverse and complex evolutionary pathways. An important factor in establishing this broad distribution is likely to be the interrelationships between the angular momentum of the galaxy and its DM halo with stellar morphology and galaxy size \citep[e.g.][]{Zavala2016,Grand2017,Dillamore2022,Lagos2022,Loubser2022,Cadiou2022StellarAM,RodriguezGomez2022,Ma2024}. Mergers can be a significant factor in determining the angular momentum of galaxies; however their angular momentum deposition is dependent in a complex way on factors such as merger mass ratios, orbital configurations, and gas content \citep[e.g.][]{Cox2006,Oogi2013,Naab2014,Choi2017,Desmond2017,Grand2017,Yoon2017,Penoyre2017,Loubser2022,Yoon2022,Jung2022,Graham2023,Lohmann2023}.

Rather than focus on the impact of mergers on angular momentum, and the impact of angular momentum on galaxy size, it is also interesting to look more directly at the impact of mergers on galaxy size. On short timescales, there is evidence that post-starburst galaxies, which in many cases (though not all) are expected to be the result of galaxy mergers, tend to be more compact compared to regular galaxies at intermediate-high redshifts (e.g. \citealt{Whitaker2012,Yano2016,Almaini2017,Wu2018,Chen2022,Setton2022}, although see \citealt{Cheng2024} who find mixed results). Over longer timescales however, it remains an open question whether mergers directly affect sizes. 

Simulating the relationship between merger histories and galaxy properties poses two major challenges: a) there is a vast range of possible merger histories, and b) the effect of any given merger history is contingent on a range of complex galaxy formation physics which cannot all be resolved from first principles, even in today's most advanced simulations. The first of these challenges -- scanning over a suitable range of merger histories -- traditionally requires a large-volume cosmological simulation to enable correlations between galaxies and their merger histories to be studied. However, with this approach, it can be hard to identify causal relationships, since every individual simulated galaxy has different formation times, merger ratios and timings, merger configurations, gas and stellar content of the merging galaxies, and large-scale environment to name but a few. The genetic modification (GM) approach \citep{Roth2016,Rey2018,Stopyra2021} instead performs controlled modifications to a set of simulation initial conditions (ICs) for one or more galaxies, targeting specific properties (e.g. a merger ratio) while constraining others (e.g. $z=0$ mass). The ICs remain consistent with a $\Lambda$CDM cosmology and are altered to the minimum extent necessary to achieve the goals. The VINTERGATAN-GM suite of zoom-in cosmological hydrodynamical simulations \citep{Rey2022,Rey2023} uses this approach to modify the early merger history of a MW-mass halo, specifically targeting the mass ratio of a \targetMerger{} merger, while maintaining the $z=0$ halo mass of the system. The simulations employ an updated version of the VINTERGATAN model \citep{Agertz2021}. The suite is thus designed to study the impact of the target merger event. 

The other challenge mentioned above is the impossibility of resolving all galaxy formation physics from first principles. In practical terms, this translates into wide variations between different simulation codes and galaxy formation subgrid recipes. Evaluation of, and comparison between, different codes are essential to making sense of this situation, and have taken two main approaches. The first is to compare the characteristics of the galaxy population as a whole i.e. with well-established scaling relations like the stellar mass-halo mass relation \citep[e.g.][]{Guo2010,Moster2013,Behroozi2019}, the size-mass relation \citep[e.g.][]{Shen2003,Lange2015}, the main sequence of star-forming galaxies \citep[e.g.][]{Brinchmann2004,Noeske2007,Speagle2014,Santini2017}, the mass-metallicity relation \citep[e.g.][]{Tremonti2004,Andrews2013,Sanders2021} and others. While most codes are calibrated to reproduce one or more of these relations, others are predictions and therefore serve as a comparative evaluation of the various simulation codes.

The second approach is to compare individual galaxies by simulating the same set of ICs with different simulation codes. This allows us to understand how the detailed differences in the way specific physical processes are implemented in the codes can lead to emergent differences in the galaxies they simulate. The Aquila project \citep{Scannapieco2012} compared the performance of nine different codes in simulating an isolated MW-mass halo. They found that despite having similar halo mass assembly histories, the simulations differed considerably in several aspects such as stellar kinematics, rotation curves, stellar and gas mass, SF history (SFH) and more, due mainly to their SF and feedback prescriptions. Similarly, the AGORA collaboration \citep{Kim2014Agora,RocaFabrega2021Agora} brings together several simulation codes (and their respective communities) to simulate the same set of ICs for a MW-mass halo. In addition to common ICs, the project attempts to homogenise the physical processes included through recalibration of tunable parameters in the codes. Following this process, a number of simulations codes converge on properties of the central galaxy, but show significant differences in e.g. the properties of the gaseous circumgalactic medium \citep{Kim2016,RocaFabrega2021Agora}. They also find that galaxy sizes (half-mass radii) can differ by nearly 1 dex between different codes evolving the same ICs \citep{RocaFabrega2024}. While powerful, such an approach relies on iterative convergence on the physical processes implemented in each of the codes in order to bring them into as close agreement as possible, and therefore does not reflect the level of uncertainty in galaxy formation physics as a whole.

Genetically modified ICs afford a third approach to comparing galaxy formation models, complementary to the two discussed above. By simulating a set of GM ICs with codes that take different approaches to galaxy formation physics, we may examine the response of the different models to the controlled variations of merger histories. Such comparisons provide insight into the impact of subgrid physics and its interaction with merger history in a way that brings a complementary perspective. We do not attempt to bring the simulation codes into agreement through any re-calibration, relying solely on the well-established codes from their respective collaborations, thereby probing the uncertainty in the physics as already implemented. The Use of GM simulations allows us to consider the origin of trends within the galaxy population in a computationally efficient way.

Following this approach, we present the PARADIGM (Probing the Altered Response of Algorithms to Diverse ICs with Genetic Modifications) project. We simulate the same set of GM ICs used for the VINTERGATAN-GM simulations \citep{Rey2023} using the IllustrisTNG model \citep{TNGMethodsWeinberger2017,TNGMethodsPillepich2018}, to produce the IllustrisTNG-GM suite of simulations. With these two suites, we explore how the two codes respond to the modifications to the ICs, thus learning about how mergers shape the properties of the galaxies they produce. While IllustrisTNG is designed and tuned to successfully reproduce known scaling relations using an effective two-phase prescription for the interstellar medium, VINTERGATAN aims to resolve the cold interstellar medium (ISM) down to temperatures of 10~K, thus allowing it to more directly model the environments where star-formation and feedback processes occur, without further tuning. As such, these two codes represent two highly successful approaches to galaxy formation simulations but with very different implementation emphases, motivating a detailed comparison.

In this paper specifically, we focus on how modified merger histories can impact galaxy sizes by altering their kinematic structure, to shed light on the long-term impacts of mergers on galaxy properties. The paper is structured as follows: Section \ref{sec:methods} provides details of the simulation suites and brief descriptions of the salient features of the two simulation codes, as well as any post-processing conducted on the simulations. In Section \ref{sec:fiducialComp}, we first discuss some comparisons of the fiducial simulations from both suites; although it is not the aim of the project to conduct one-to-one comparisons of the simulated galaxies, we discuss some key differences in this section to establish the context in which to interpret the subsequent results. In Section \ref{sec:sizeComp}, we first examine the connection between $z=0$ galaxy sizes and merger histories, along with the evolution in galaxy sizes. In Section \ref{sec:kinComp}, we then explore how the kinematic structure of galaxies depends on changes to the merger histories in order to understand the trends in galaxy sizes. Finally, we discuss our results in Section \ref{sec:discussion} and present our conclusions in Section \ref{sec:conclusions}.


\section{Simulations and methods} \label{sec:methods}

\subsection{Common simulation setup}

\begin{table*}
    \caption{Key properties of the $z=2$ merger that is targeted when modifying the ICs for both sets of simulations and the two additional realisations of the \tng{} \XC{} simulation, to illustrate the level of stochastic uncertainty in these properties. The start of a merger is defined as the time of accretion of the secondary galaxy i.e. the last time the secondary galaxy was outside the virial radius of the main galaxy's FOF halo. The end of the merger is defined as the time of coalescence i.e. when the secondary galaxy can no longer be identified as separate from the main galaxy. The halo mass, stellar mass, and $v_{\rm max}$ ratios are calculated at the time of accretion (note that these values are taken directly from the halo finders and therefore may not be directly comparable between the two simulation codes). $v_{r}$ and $v_{\theta}$ are the radial and tangential components of velocity of the secondary galaxy relative to our main galaxy at the the start of the merger i.e. just prior to accretion. We also provide the impact parameter, $b$, measured at this time.} \label{tab:gmProperties}
    \centering
    \begin{tabular}{|l|c|c|c|c|c|c|c|c|c|c|c|c|c|c|c|}
        \hline
         &  & \multicolumn{5}{c}{VINTERGATAN-GM} &  & \multicolumn{8}{c}{IllustrisTNG-GM}\\
        \hline
         &  & \XXC{} & \XVC{} & \XC{} & \XCX{} & \XCXX{} &  & \XXC{} & \XVC{} & \XC{} & \XCX{} & \XCXX{} & & \XCRII{} & \XCRIII{} \\
        \hline
        $z_{\rm start,z=2\ merger}$ &  & 1.90 & 1.99 & 1.99 & 1.94 & 2.23 &  & 2.02 & 2.02 & 2.05 & 1.96 & 2.26 &  & 2.05 & 2.05 \\ 
        $z_{\rm end,z=2\ merger}$ &  & 1.70 & 1.56 & 1.79 & 1.51 & 1.78 &  & 1.84 & 1.81 & 1.84 & 1.76 & 2.02 &  & 1.84 & 1.84 \\ 
        $t_{\rm start,z=2\ merger}$ [Gyr] &  & 3.46 & 3.30 & 3.30 & 3.38 & 2.96 &  & 3.25 & 3.25 & 3.21 & 3.36 & 2.91 &  & 3.21 & 3.21 \\ 
        $t_{\rm end,z=2\ merger}$ [Gyr] &  & 3.83 & 4.13 & 3.65 & 4.26 & 3.68 &  & 3.56 & 3.61 & 3.56 & 3.71 & 3.25 &  & 3.56 & 3.56 \\
        Halo mass ratio &  & 1:10.6 & 1:9.7 & 1:6.0 & 1:3.5 & 1:2.2 &  & 1:18.5 & 1:19.4 & 1:13.0 & 1:7.0 & 1:3.5 &  & 1:12.2 & 1:14.0 \\ 
        Stellar mass ratio &  & 1:24.6 & 1:15.7 & 1:8.2 & 1:4.7 & 1:2.1 &  & 1:51.9 & 1:36.9 & 1:20.9 & 1:8.4 & 1:2.6 &  & 1:20.7 & 1:20.8 \\
        $v_{\rm max}$ ratio &  & 1:4.2 & 1:2.4 & 1:2.1 & 1:1.9 & 1:2.6 &  & 1:4.0 & 1:3.9 & 1:3.9 & 1:2.9 & 1:2.6 &  & 1:3.8 & 1:3.7\\
        $v_{r}(t_{\rm start})$ [km/s] &  & 133 & 130 & 114 & 119 & 125 & & 141 & 138 & 117 & 144 & 139 &  & 119 & 116 \\
        $v_{r}/v_{\theta}(t_{\rm start})$ &  & 9.3 & 9.0 & 15.8 & 6.8 & 5.5 & & 7.6 & 8.5 & 17.2 & 7.8 & 8.3 &  & 18.7 & 19.5\\
        $b(t_{\rm start})$ [kpc] &  & 7.8 & 7.9 & 4.1 & 10.9 & 13.4 & & 9.7 & 8.7 & 4.1 & 9.8 & 8.7 &  & 3.7 & 3.6\\
        \hline
    \end{tabular}
\end{table*}

This study makes use of two suites of zoom-in cosmological hydrodynamical simulations of MW-mass haloes, VINTERGATAN-GM and IllustrisTNG-GM, as part of the PARADIGM project. Each suite evolves the same set of five ICs, a fiducial (un-modified) IC for a MW-mass halo at $z=0$, and four `genetically modified' (GM) versions which modify the strength of a significant merger that the galaxy experiences at \targetMerger{}, while maintaining its $z=0$ halo mass. The target halo in the fiducial ICs was chosen from an initial DM-only, uniform-resolution simulation of a $(50 h^{-1} {\rm Mpc})^{3}$ volume with $512^{3}$ particles and a mass resolution of $m_{\rm DM}=1.2\times10^{8}\MSUN$. The halo was selected to have $\MHOST \approx 10^{12}\MSUN$ at $z=0$ and to be isolated with no massive neighbours within $5\,\RHOST$. Additionally, it has an active merger history until \targetMerger{} and a relatively quiet history at later times. The zoom region refines particles within $3\,\RHOST$ at $z=0$ to have a DM particle mass of $2.0\times 10^{5}\MSUN$ and an initial minimum gas cell mass of $3.7\times 10^{4}\MSUN$. For all simulations, we assume a flat $\Lambda$CDM cosmology with cosmological parameters $\Omega_{\rm m,0}=0.3139$, $h=0.6727$, $n_{s}=0.9645$ and $\sigma_{8}=0.8440$ \citep{Planck2016}.

All the ICs were generated with the code \textsc{GenetIC} \citep{Roth2016,Rey2018,Stopyra2021}, which allows us to perform the four controlled modifications to the ICs. In particular, by altering the average overdensity of the region that the halo is embedded in at early times ($z=99$) to 90, 95, 110 and 120 per cent of its fiducial value, we alter the merger mass ratio of the targeted merger at \targetMerger{}. For complete details regarding the generation of the ICs, see \citet{Rey2022} and \citet{Rey2023}, where the VINTERGATAN-GM project is also introduced. For the remainder of the text of this paper, we refer to the fiducial ICs (and the resultant simulations) as \XC{} and the four GM ICs as \XXC{}, \XVC{}, \XCX{} and \XCXX{}, corresponding to the smallest, smaller, larger and largest \targetMerger{} merger. Table \ref{tab:gmProperties} provides some key properties of the targeted mergers from the two suites of simulations.  

The ICs were evolved with two different simulation codes: the VINTERGATAN model \citep{Agertz2021} and the IllustrisTNG model \citep{TNGMethodsWeinberger2017,TNGMethodsPillepich2018}. We outline the details of each model and in the following sections. It should be emphasized that neither of the two codes were re-calibrated for the present study; rather, we adopted the parameter configurations for the VINTERGATAN model from \citet{Agertz2021} and specifically for the TNG50 simulations from \cite{TNG50Pillepich2019,TNG50Nelson2019}, the latter of which has been tuned to reproduce key established $z=0$ scaling relations. 

In addition to the above two sets of simulations, we also performed two additional simulations with the fiducial ICs using the IllustrisTNG code (referred to as \XCRII{} and \XCRIII{}), in order to quantify the scatter associated with stochasticity in this model. Since the VINTERGATAN-GM simulations are computationally expensive, we did not attempt a similar test for the VINTERGATAN-GM suite, although such stochastic uncertainty certainly exists for this model as well. For the remainder of the paper, we will refer to the IllustrisTNG and VINTERGATAN models and the corresponding sets of simulations as \tng{} and \vgn{} respectively.

\subsection{\vgn{} simulations}
The \vgn{} simulations were generated with the Adaptive Mesh-Refinement (AMR) code \textsc{ramses} \citep{Teyssier2002}, which employs a particle-mesh algorithm to solve Poisson's equation and an HLCC Riemann solver \citep{Toro1994} for fluid dynamics assuming an ideal equation of state for gas with $\gamma=5/3$. The AMR strategy within \textsc{ramses} allows us to reach a minimum effective gas cell radius of 11~pc (and a median value of 88~pc) throughout the central galaxy at $z=0$ in the fiducial simulation. The simulations are based on the VG galaxy formation model \citep{Agertz2021}, including prescriptions for SF, feedback from supernovae (SNe), metal enrichment, gas cooling through metal-line emission, and heating and photo-ionization from a uniform, time-dependent UV background; we summarise the salient features here. Note that the DM and baryon mass resolution of our current simulations are a factor of $\sim6$ lower than those of the original VINTERGATAN simulations.

SF occurs in gas cells when its gas density exceeds $100\,m_{\rm p}\,cm^{-3}$ and its temperature is below 100~K, at which time stars are formed through a Poisson process following a Schmidt law \citep{Schmidt1959}, with new stellar particles having a stellar mass of $10^{4}\MSUN$ and modelled as single stellar populations, adopting a \citet{Chabrier2003} initial mass function (IMF). The SF efficiency (SFE) $\epsilon_{\rm ff}$ is set to 10 per cent following \citet{Agertz2020}. The model accounts for SNeIa and SNeII and stellar winds from O, B and AGB stars in determining the injection of mass, energy, momentum and metals from each stellar particle, dependent on its age, mass and metallicity \citep[see][for details]{Agertz2013}. Feedback from SNe is modelled following the \citep{Kim2015} prescription, in the form of energy injected into the gas cell nearest to the star particle when the cooling radius for the SN explosion is resolved by by at least six gas cells, and otherwise in the form of momentum into the cells surrounding the star particle. The model also tracks the abundances of iron (Fe) and oxygen (O) separately, with yields from SNe taken from \citep{Woosley2007}. Note that there is no prescription for active galactic nuclei (AGN) in the model.

\paragraph*{Halo-finding:} We identify bound haloes and the galaxies embedded within using \textsc{ahf} \citep[the AMIGA Halo Finder][]{Gill2004,Knollmann2009}, which also uses an AMR strategy to identify the peak overdensities in the DM and baryonic mass distribution. The corresponding haloes at such overdensities are then constructed following an iterative unbinding procedure to isolate bound haloes. 

\subsection{\tng{} simulations}
The \tng{} simulations were performed using the moving-mesh code \textsc{arepo} \citep{Springel2010} which solves the equations of magneto-hydrodynamics (MHD) following a finite volume method on a Voronoi-tessellated, unstuctured moving mesh. The generating points for the Voronoi mesh are allowed to move at close to the local fluid velocity. The minimum effective cell radius at $z=0$ attained in the fiducial simulations is 46~pc (with a median value of 174~pc). The simulations employ the TNG galaxy formation model, specifically the parameters used for the TNG50 suite of simulations \citep{TNGMethodsWeinberger2017,TNGPillepich2018,TNG50Nelson2019,TNG50Pillepich2019}, including prescriptions for SF, formation of super-massive BHs (SMBHs), stellar feedback, feedback from AGN, metal enrichment, gas cooling through metal-line emission, and heating though a uniform, time-dependent UV background; we summarise the key ingredients here. 

SF is modelled following the \citet{Springel2003} model whereby gas cells above a density threshold of $0.1\,m_{p}\,cm^{3}$ and below a variable temperature threshold are allowed to form stars in a stochastic manner, following a Kennicutt-Schmidt relation \citep{Schmidt1959,Kennicutt1998} and assuming a \citet{Chabrier2003} IMF. To account for the limited resolution in the ISM available in the large-volume simulations in the IllustrisTNG suite, the ISM is treated as a two-phase model with an effective equation of state (eEOS), and an eEOS parameter of $q_{\rm EOS}=0.3$ \citep[see][for details]{Vogelsberger2013,Torrey2014}. The temperature threshold is also determined by the eEOS, up to a minimum of $10^{4}$~K. The model accounts for the return of mass and metals from SNeIa and SNeII and AGB stars. Stellar feedback is modelled in the form of wind particles, which carry mass, energy, momentum and metals, based on the available SN energy. The wind particles are effectively hydrodynamically uncoupled until they leave the local ISM, after which they are recoupled and can deposit mass, energy, momentum and metals back into the galaxy. SMBHs are seeded in haloes with a minimum mass of $5\times10^{10}\ h^{-1}\MSUN$ and have an initial mass of $8\times10^{5}\ h^{-1}\MSUN$. They accrete surrounding gas through Bondi accretion and return energy to their surroundings as AGN feedback. The feedback models consists of two phases depending on their accretion rate (which itself is dependent on the BH mass and the density and sound speed of its surrounding gas): in the `high-accretion' mode, AGN feedback is in the form of thermal energy injected continuously into the surrounding gas cells; in the `low-accretion' mode, it takes the form of a momentum injection, whereby momentum is stored up and released stochastically into the surrounding gas cells. The code tracks the abundance of H, He, C, N, O, Ne, Mg, Fe, Si separately, with yields taken from several different sources \citep[see][for details]{TNGMethodsPillepich2018}.

\paragraph*{Halo-finding:} Bound haloes and galaxies were identified within the simulations in a two step process, identically to the methodology of the IllustrisTNG suite of simulations. First, haloes are detected with a Friends-of-Friends (FOF) algorithm \citep{Davis1985} applied to the (high-resolution) DM particles, with a linking length of 0.2 times the mean inter-particle distance. Galaxies are then identified as bound structures within each FOF group using the \textsc{subfind} algorithm \citep{Springel2001,Dolag2009} applied to both DM and baryonic elements.

\subsection{Common post-processing} \label{sec:postprocessing}

We rely on the individual halo finders mentioned above to identify bound haloes and the galaxies embedded within them. However, since there are significant differences in the methods employed by the two halo finders, especially with regards to identifying substructure, it is expected that they may return different results for many galaxy properties. Hence, we rely on the halo finders solely to locate the positions of the main halo and to determine which resolution elements are associated with it; most other properties are then measured by common post-processing methods as detailed below. As such, we do not expect any differences in the halo finders to significantly affect our results. We make extensive use of the \textsc{pynbody} \citep{Pontzen2013} and \textsc{tangos} \citep{Pontzen2018} packages to calculate all the properties used in this study, as described below.

\paragraph*{Merger trees:} We employ a simple merger tree algorithm within \textsc{tangos}, whereby haloes are connected to their descendant haloes (those containing the majority of the current DM particles) in the next timestep. The main progenitor is chosen to be the halo with the largest contribution of DM particles when travelling the tree backwards through cosmic time.

\paragraph*{Luminosities/magnitudes:} Luminosities for the stellar particles are calculated by interpolating from Padova isochrones for simple stellar populations (SSPs) on a grid of ages and metallicities \citep{Marigo2008,Girardi2010}, where the ages and metallicities for the stellar particles are retrieved directly from the respective simulations.

\paragraph*{Sizes:} In order to separate the central galaxy from its substructure and to determine the extent of the galaxy in a consistent and observationally motivated manner, we first define the edge of the galaxy based on its surface brightness profile in a face-on orientation. The maximum radius $\RSB$ is taken to be the radius at which the V-band surface brightness equals 26.5~mag/arcsec$^{2}$, consistent with the sensitivity of several observational surveys, although we have confirmed that our results are not significantly different if using a fainter limit of 28~mag/arcsec$^{2}$. The size of the galaxy is then characterised as a (3D) half-mass radius $\RHALF$ and a (2D) face-on half-light radius $\RHALFV$ within this maximum extent. Other properties are calculated with resolution elements within a multiple of $\RHALF$ unless specified otherwise. Note that in measuring masses and other galaxy properties, we used radial profiles of the properties and interpolated values within a given radial aperture. Since we adopted log binning for the profiles, we often show quantities measured within $10^{0.5} \approx 3.2\,\RHALF$, but simply refer to this as $3\,\RHALF$ for brevity. We discuss measurement methods in more detail in Section \ref{sec:fiducialComp}.

\paragraph*{Circularity parameter:} To characterise the kinematics of the galaxies, we follow the methodology of \citet{Marinacci2014,Genel2015} and define a circularity parameter for each resolution element $\epsilon = j_{z}/j_{\rm max}(E)$, where $j_{z}$ is the angular momentum of the element parallel to the disc of the galaxy and $j_{\rm max}(E)$ is the maximum possible angular momentum for a given total kinetic energy. In doing so, we determine the plane of the galaxy disc using stellar particles within 5 kpc of its centre. Following the methodology of \citet{Genel2015}, $j_{\rm max}(E)$ is taken to be the maximum $j_{z}$ of the 50 preceding and following particles when ordered by their total kinetic energy. \footnote{We have confirmed that increasing the number of particles to 100 or 500 does not have a noticeable impact on the circularity parameter distributions.}


\section{Comparison of the fiducial \tng{} and \vgn{} galaxies} \label{sec:fiducialComp}

\begin{figure*}
    \includegraphics[width=\linewidth]{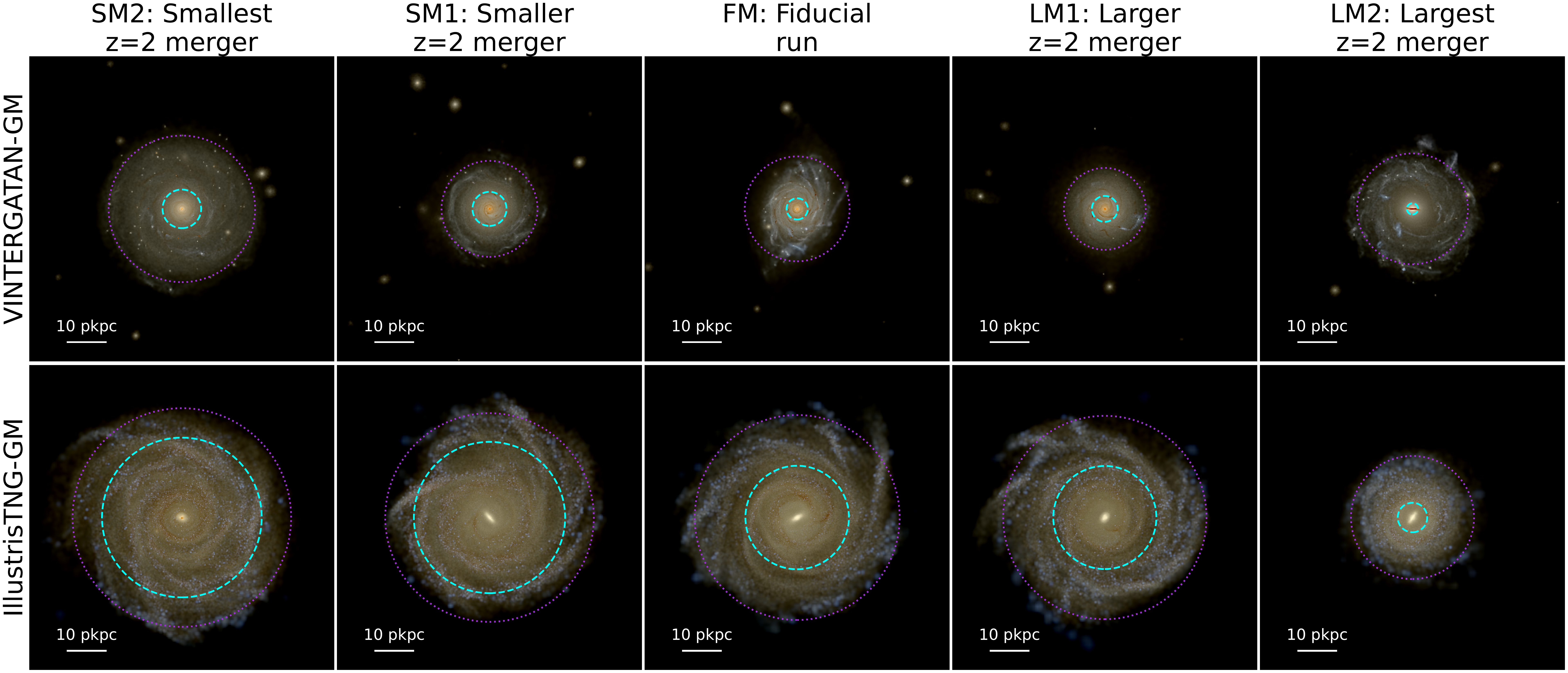}
    \includegraphics[width=\linewidth]{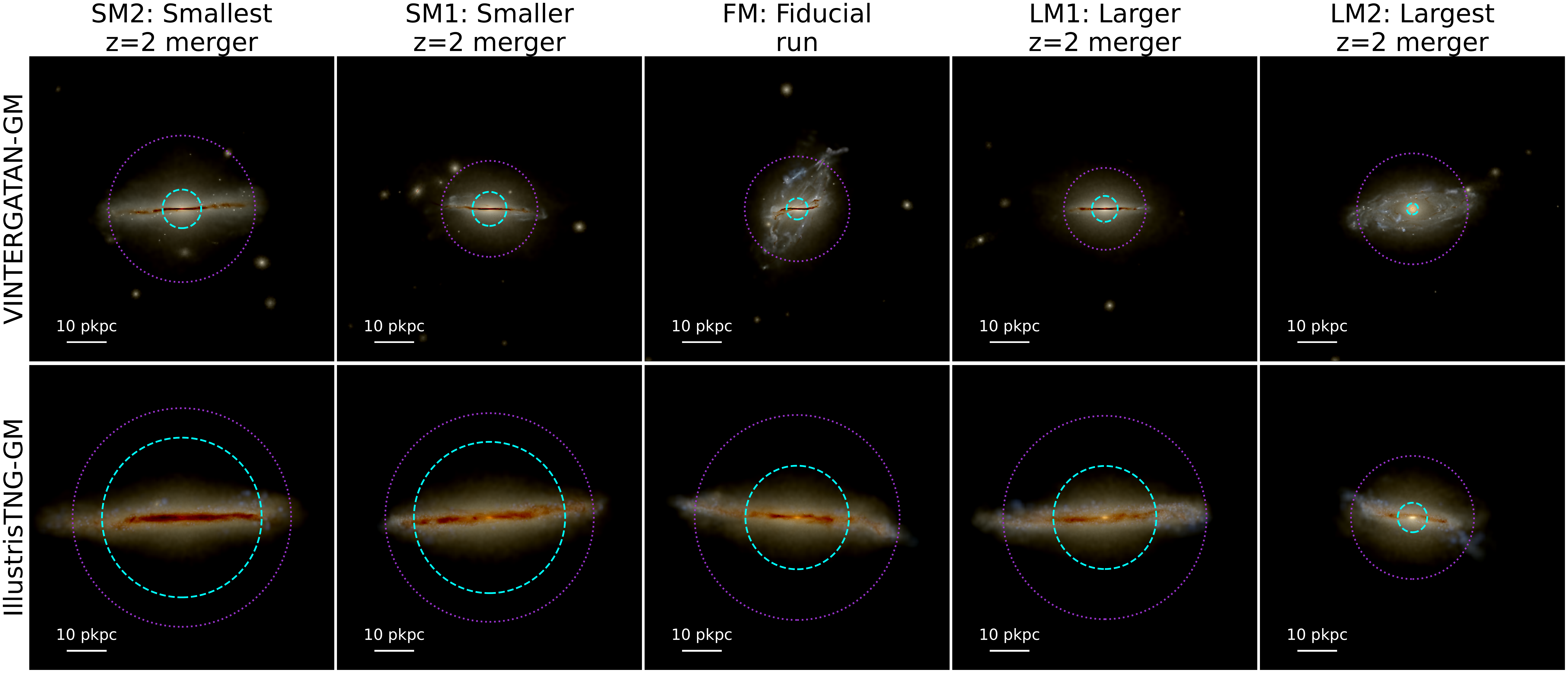}
    \caption{Mock images of the galaxies in our sample in face-on orientation (top rows) and edge-on orientation (bottom rows). Colours correspond to Johnsons U, V and I bands, and cover 18-28~mag/arcsec$^{2}$ in surface brightness. Each image is $80 \times 80$~kpc in size. The minimum spatial resolution (measured as the minimum effective gas cell radius) is 46~pc and 11~pc for the \tng{} and \vgn{} galaxies respectively, significantly smaller than the pixel size of 200~pc in the images. We have applied a simple correction for dust absorption following the \citet{Calzetti2000} law using routines from the \textsc{extinction} package \citep{extinctionPackage} and \textsc{pynbody}. The purple dotted circle indicates the maximum radius, $\RSB$, where $\mu_{\rm V-band}=26.5$~mag/arcsec$^{2}$, while the cyan dashed circle indicates $3\times$ the stellar half-mass radius $\RHALF$. There is a significant correlation apparent between decreasing galaxy sizes -- as measured by $\RHALF$ -- and increasing \targetMerger{} merger mass ratio. The trend in $\RSB$ is less clear, especially in the \vgn{} case, since this radius is highly sensitive to the presence of non-uniformly distributed young stars in the outskirts of the galaxies.} \label{fig:stellarMocks}
\end{figure*}

\begin{table}
    \caption{Properties of the central galaxies at $z=0$ in the \XC{} simulations from each code.}  \label{tab:fiducialProperties}
    \centering
    \begin{tabular}{l|c|c}
        \hline
         & VG & TNG \\
        \hline
        $R_{\rm 200c}$ [kpc] & 201.8 & 208.2 \\
        $M_{\rm 200c}\ [10^{10}\MSUN]$ & 86.5 & 95.0 \\
        $M_{\rm 200c,*}\ [10^{10}\MSUN]$ & 1.8 & 5.2 \\
        $M_{\rm 200c,gas}\ [10^{10}\MSUN]$ & 2.6 & 4.7 \\
        $f_{\rm baryon,200c}$ & 5.2\% & 10.4\% \\
        $f_{\rm *,200c}$ & 2.1\% & 5.5\% \\
        $f_{\rm gas,200c}$ & 3.0\% & 4.9\% \\
        \hline
        $r_{\rm SB=26.5,V-band}$ [kpc] & 13.8 & 26.9 \\
        $r_{\rm *,1/2,V-band}$ [kpc] & 2.04 & 9.88 \\
        $r_{\rm *,1/2,mass,3D}$ [kpc] & 0.93 & 4.53 \\
        \hline
        $M_{\rm *}(<3\,\RHALF)\ [10^{10}\MSUN]$ & 1.12 & 4.44 \\
        $M_{\rm gas}(<3\,\RHALF)\ [10^{10}\MSUN]$ & 0.11 & 0.29 \\
        $M_{\rm *}(<10\,\RHALF)\ [10^{10}\MSUN]$ & 1.46 & 5.08 \\
        $M_{\rm gas}(<10\,\RHALF)\ [10^{10}\MSUN]$ & 0.45 & 1.42 \\
        \hline
    \end{tabular}
\end{table}

\begin{figure}
    \centering
    \includegraphics[width=0.95\linewidth]{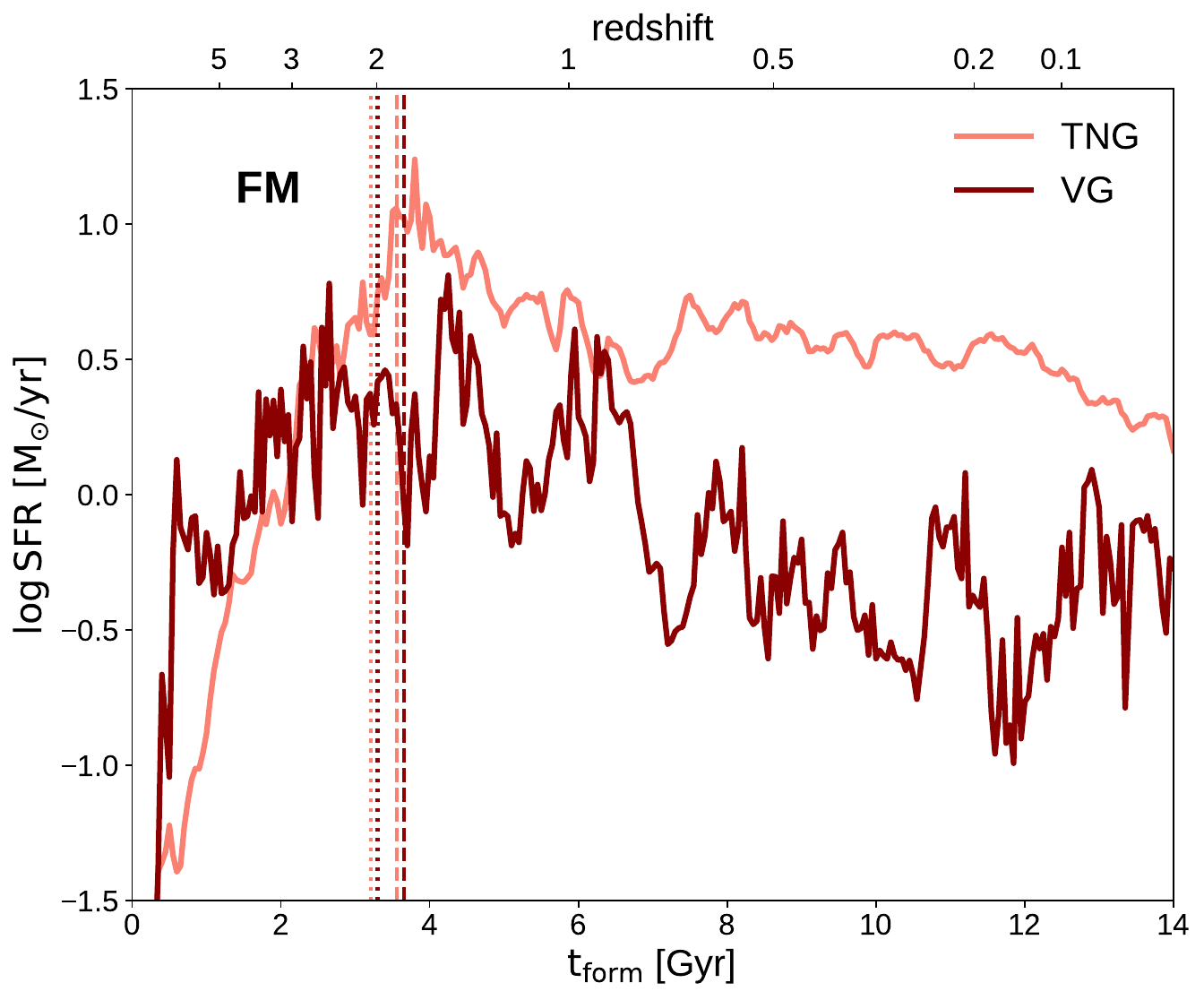}
    \caption{Instantaneous SFR as a function of time for the \XC{} galaxies, measured using the ages of the stellar particles within $\RSB$ (in 3D) at $z=0$. Note that we do not differentiate between in-situ and ex-situ stars. The TNG galaxy initially has a lower SFR compared to the VG galaxy until $z \sim 3$. However, before and during the \targetMerger{} it has a significantly higher SFR, and then continues to have an SFR greater than the VG galaxy by up to 1 dex until present day.} \label{fig:fiducialSFR}
\end{figure}

\begin{figure}
    \centering
    \includegraphics[width=0.95\linewidth]{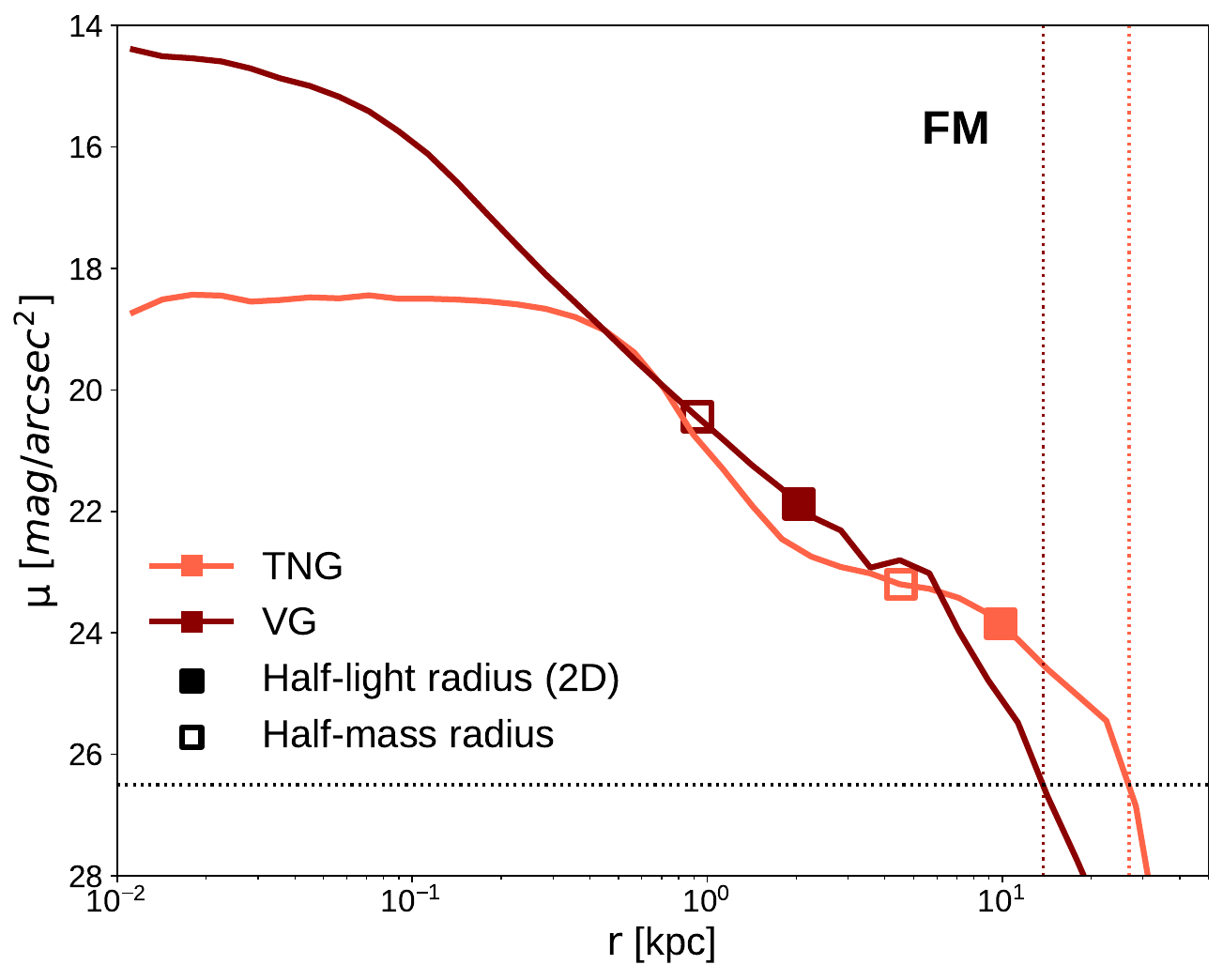}
    \caption{Surface brightness profile in V-band for the \XC{} galaxies. The horizontal dotted line demarcates $\mu_{\rm V}=26.5$ mag/arcsec$^{2}$, our chosen threshold to define the edge of the galaxy. Vertical dotted lines indicate the resulting sizes ($\RSB$) for the two galaxies. The corresponding filled symbols indicate the stellar 2D half-light radius, while the open symbols indicate the stellar 3D half-mass radius. The TNG surface brightness profile is significantly flatter and fainter within the inner $\sim 200$ pc compared to the VG profile, but brighter in the outer regions of the galaxies, i.e. beyond $\sim 5$ kpc. Thus, the $\RSB$ for the TNG galaxy is twice as large as the VG galaxy, while the half-light/mass radii differ by a factor of nearly 5.} \label{fig:fiducialProfiles}
\end{figure}

The aim of this study is to examine the similarities and differences in the correlations between galaxy properties and merger histories as predicted by different simulations codes. In order to establish the context within which to interpret later trends, in this section we provide initial comparisons between the galaxies from both sets of simulations, focusing on the \XC{} case.

Fig. \ref{fig:stellarMocks} shows mock images of all the galaxies in our suite, with the two fiducial galaxies in the middle columns. The red, green and blue colours correspond to Johnsons U, V and I bands respectively, where the luminosities in each band were measured as described in Section \ref{sec:postprocessing}. In making the images, using \textsc{pynbody}, we first generated a stellar emission map in each band. The maps can then be scaled as required (here we choose to scale I band luminosities by a factor of 0.5 to better match visual colours), and normalised to the chosen surface-brightness range. These normalised maps are combined to generate the final image. Importantly, all simulations in our sample are processed in the same manner. For illustration purposes, we have applied a simple correction for dust absorption (before the scaling and normalization of the raw maps in each band) assuming that dust column density is proportional to the total metal column density and following the \citet{Calzetti2000} law, using the \textsc{extinction} Python package \citep{extinctionPackage}. Note however, that we do not apply the dust correction when measuring galaxy sizes (half-light radii) as described in Section \ref{sec:postprocessing}, due to the added uncertainty in any such dust modelling and since we use half-mass radii for most of the later analyses. Finally, Table \ref{tab:fiducialProperties} also provides various properties of the fiducial central galaxies from both simulations. 

The two fiducial simulations are significantly different as seen from Fig. \ref{fig:stellarMocks} and Table \ref{tab:fiducialProperties} -- the \tng{} halo contains nearly 3 times the stellar mass and nearly twice the gas mass within the virial radius as the \vgn{} halo, resulting in a baryon fraction that is twice that in the \vgn{} simulation. This indicates that over the lifetime of the galaxy, the VG model leads to significantly less baryonic matter being retained within the halo as well as to a lower proportion of the gas being converted into stars. To understand these differences, in Fig. \ref{fig:fiducialSFR}, we show the instantaneous star-formation rates (SFRs) of the two fiducial galaxies, measured using the ages of the stellar particles within $\RSB$ (in 3D) at $z=0$. The TNG model leads to a later start to the build up of stellar mass in the galaxy and a lower SFR at early times until $z\sim3$. After this time however, the \tng{} galaxy's SFR surpasses that of the \vgn{} galaxy by nearly 1 dex, and remains so until $z=0$. This later phase of SF is therefore responsible for the \tng{} halo and galaxy containing significantly more stellar mass at $z=0$. Despite these significant differences in the central galaxy properties, the \tng{} fiducial halo's virial radius is only 3\% larger than the \vgn{} fiducial halo (and it follows that the former has a virial mass that is 10\% higher than that of the latter), confirming that the overall structure formation within each simulation is in agreement with each other. 

Table \ref{tab:gmProperties} shows that the targeted merger in the two fiducial simulations occurs at similar times and has similar infall velocities and impact parameters. There are more significant differences between the merger mass ratios, though we caution that these values rely on properties from the two halo finders and may not be directly comparable. While previous studies have shown that the mass of idealized, isolated, haloes agrees to within a few per cent between \textsc{ahf} and \textsc{subfind} \citep[][]{Knebe2011}, when considering haloes and subhaloes in a cosmological context, the differences can be as large as $\sim 40$ per cent \citep[][]{Onions2012}. The latter of these is especially relevant during mergers, but is still significantly smaller than the differences in merger mass ratios in our simulations. Instead, we have confirmed that they are due to the \tng{} haloes, both the primary and secondary in the merger, being significantly \emph{less} massive than the \vgn{} ones at these early epochs; while the primary halo is a factor of $\sim 2$ less massive, the secondary is a factor of $\sim 4$ less massive, leading to the larger mass ratio of 1:13 in the \tng{} case compared to 1:6 for the \vgn{} case. This points to differences in the growth of structure between the models at these early epochs. Thus, it appears that the \tng{} fiducial galaxy experiences a milder merger (i.e. having a smaller mass ratio) compared to the \vgn{} one. As we show later, this also has an impact on the kinematic make-up of all galaxies in our sample.

While the stellar mass within the virial radius differs by a factor of nearly 3, the stellar mass within the galaxy itself i.e. within $3\,\RHALF$ differs by a factor of nearly 4. This indicates that a higher proportion of the stellar mass within the \vgn{} halo is contributed by the stellar halo, and more importantly dwarf satellites, compared to the \tng{} halo. This can also be seen in Fig. \ref{fig:stellarMocks}, where 4-5 satellites are visible in the vicinity of the VG galaxy, while none are seen within the same spatial extent for the TNG galaxy. The two galaxies also have significantly different surface brightness profiles, not only in terms of spatial extent, but also in terms of shape, as shown in the top panel of Fig. \ref{fig:fiducialProfiles}. The TNG galaxy has a flatter and fainter surface brightness profile in the inner regions i.e. $\lesssim 200$ pc, but a brighter profile in the outer regions i.e. $\gtrsim 5$ kpc compared to the VG galaxy, as was also seen in the mock images. In the intermediate regions however, i.e. $0.5-5$ kpc, the two fiducial galaxies have similar values of surface brightness. The presence of AGN feedback in the TNG model results in the ejection of significant amounts of gas from the central regions thus preventing SF, which produces the flatter and fainter central surface brightness and a profile that can be characterised as two distinct S\'{e}rsic profiles. The VG model does not include any such mechanism and therefore allows for gas to condense to much smaller radii, resulting in a more concentrated profile that is brighter in the central regions and can be characterised as a single Sersic profile. Differences in the mass profiles of the galaxies (not shown) mean that an aperture of e.g. $3\,\RHALF$ contains differing proportions of the total mass of the galaxy (90 per cent versus 75 per cent). In later sections, we use this aperture as an observationally motivated measure of the size of the galaxy, and so highlight here that there are inherent limitations in any such single number due to the differing shapes of the mass/light profiles of the galaxies.

Previous studies of the IllustrisTNG suite of large-volume simulations have found a diversity of galaxy sizes which has been shown to be linked to the range of AM of their surrounding haloes, coupled with an efficient transfer of AM between halo and galaxy \citep{TNG50Pillepich2019,RodriguezGomez2022,Ma2024}. Additionally, they find that halo formation times also correlate with galaxy sizes: earlier forming haloes host more compact galaxies compared to later forming ones. Additionally, \citet{Wang2023} have shown that the TNG50 discs are larger than the average for observed galaxies, which they attribute to the lack of viscous processes that can help gas shed angular momentum. Given the greater computational expense of resolving denser gas, a \vgn{} volume is not currently feasible. Therefore, it is not known whether the VG model reproduces the diversity of galaxy sizes seen in observations. However, simulations in the extended VINTERGATAN-GM suite using different ICs have also produced more extended discs (Agertz et al., in prep). In the current simulations, it is possible that variations to the model result in overcooling of gas, promoting bulge growth; we defer a detailed exploration to future work. As we show in the following section, both fiducial galaxies are consistent with observational results in the size-mass plane, but represent different morphological sub-populations. In the following sections, we move past such differences in order to focus on the response of the two codes to the modified ICs.


\section{Comparison of trends due to modified ICs} \label{sec:sizeComp}

\begin{figure}
    \centering
    \includegraphics[width=0.95\linewidth]{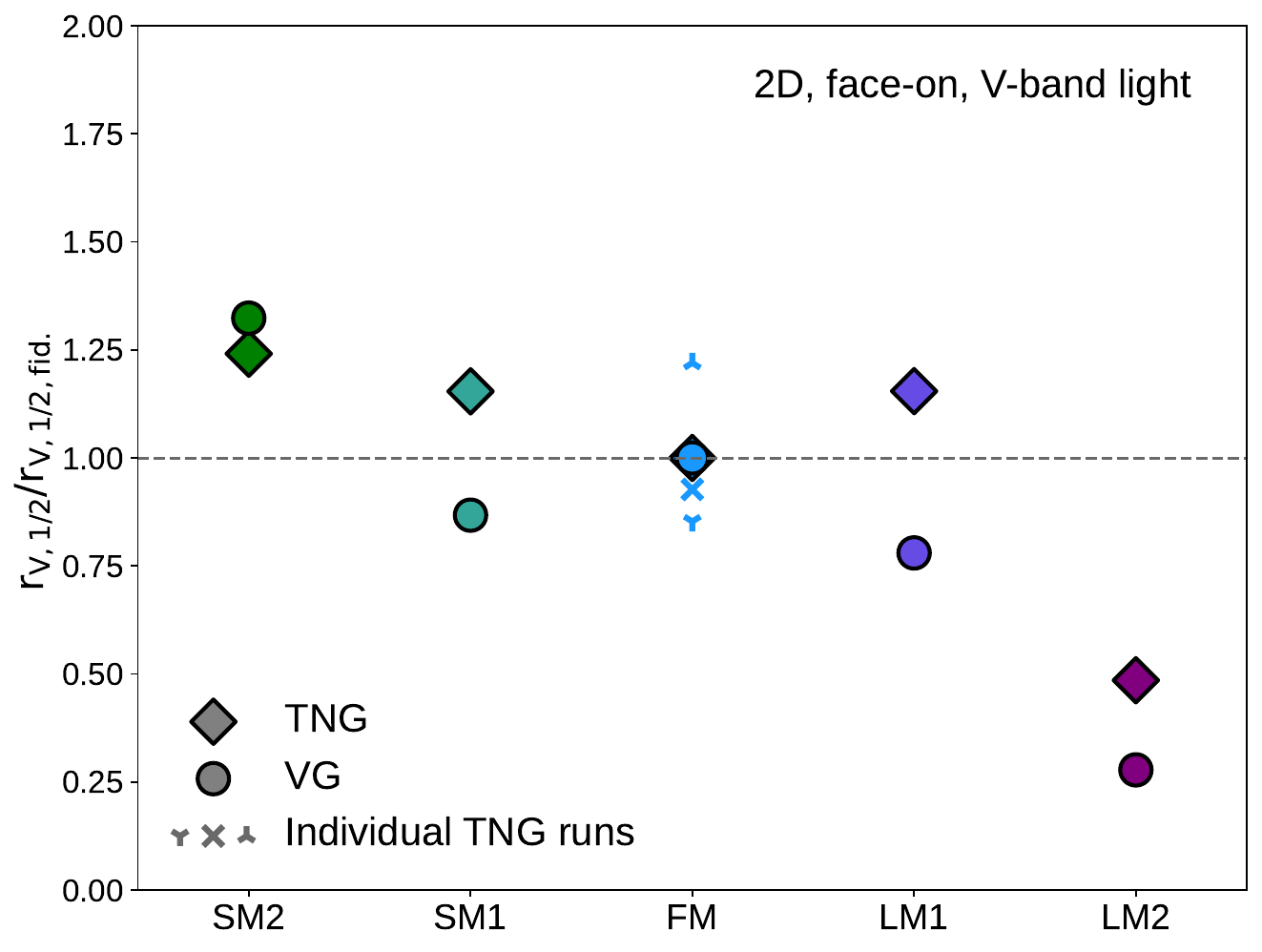}
    \includegraphics[width=0.95\linewidth]{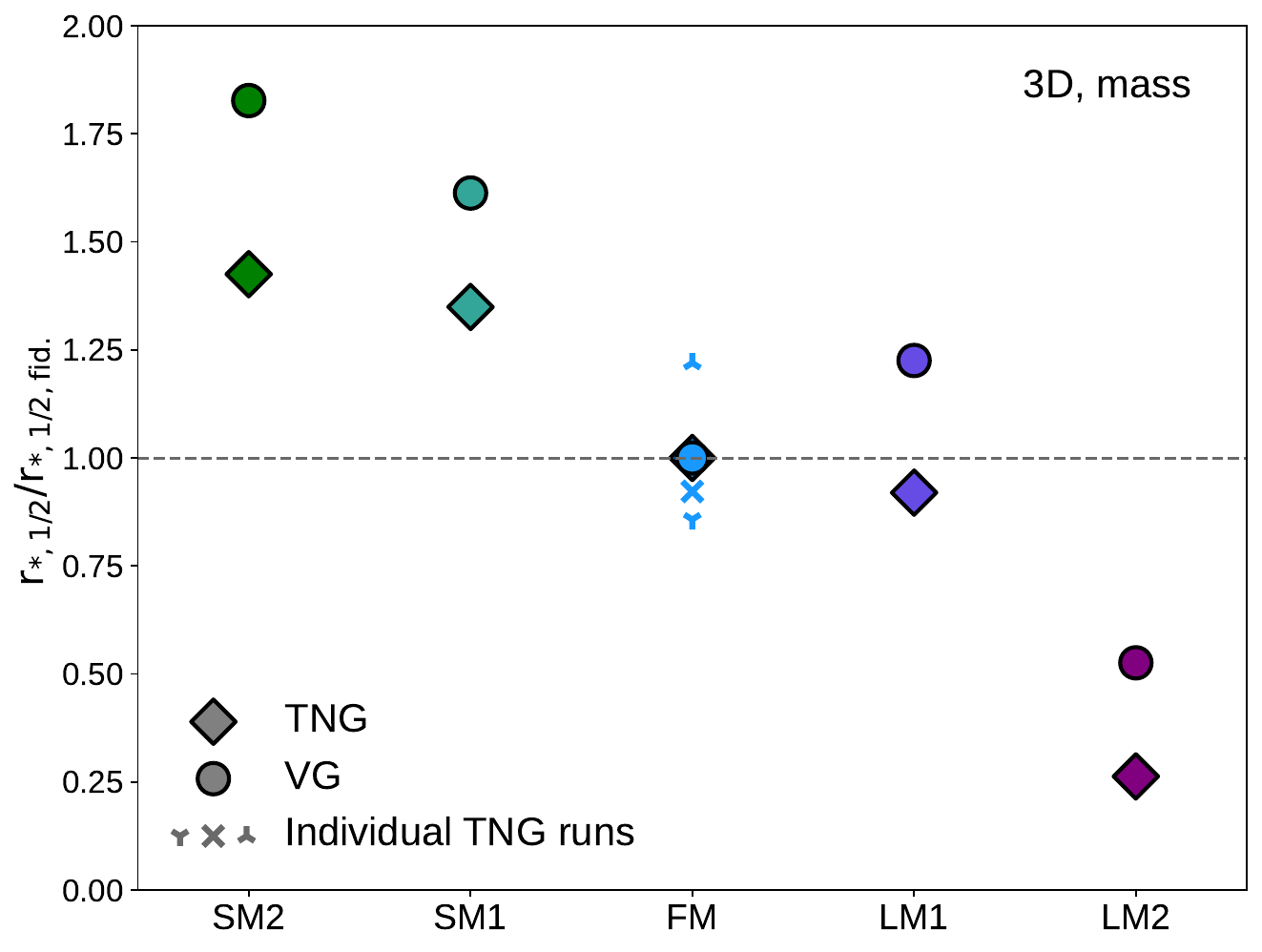}
    \caption{Half-light (top) and half-mass (bottom) radii for the centrals at $z=0$, normalised by the corresponding fiducial values from the two sets of simulations. In the case of the fiducial \tng{} simulation, we show the average of the three runs (large blue diamond) and normalise by this value, while the results for the individual runs are shown by the small cross, up and down markers. There is a clear trend of increasing $z=0$ size with decreasing \targetMerger{} merger mass ratio, albeit with some scatter. The trends are stronger in half-mass radius compared to half-light radius, but significant in both nonetheless. The radii from the smallest/largest merger scenarios can differ by as much as 80 per cent relative to the fiducial value.} \label{fig:sizesToday}
\end{figure}

\begin{figure}
    \centering
    \includegraphics[width=\linewidth]{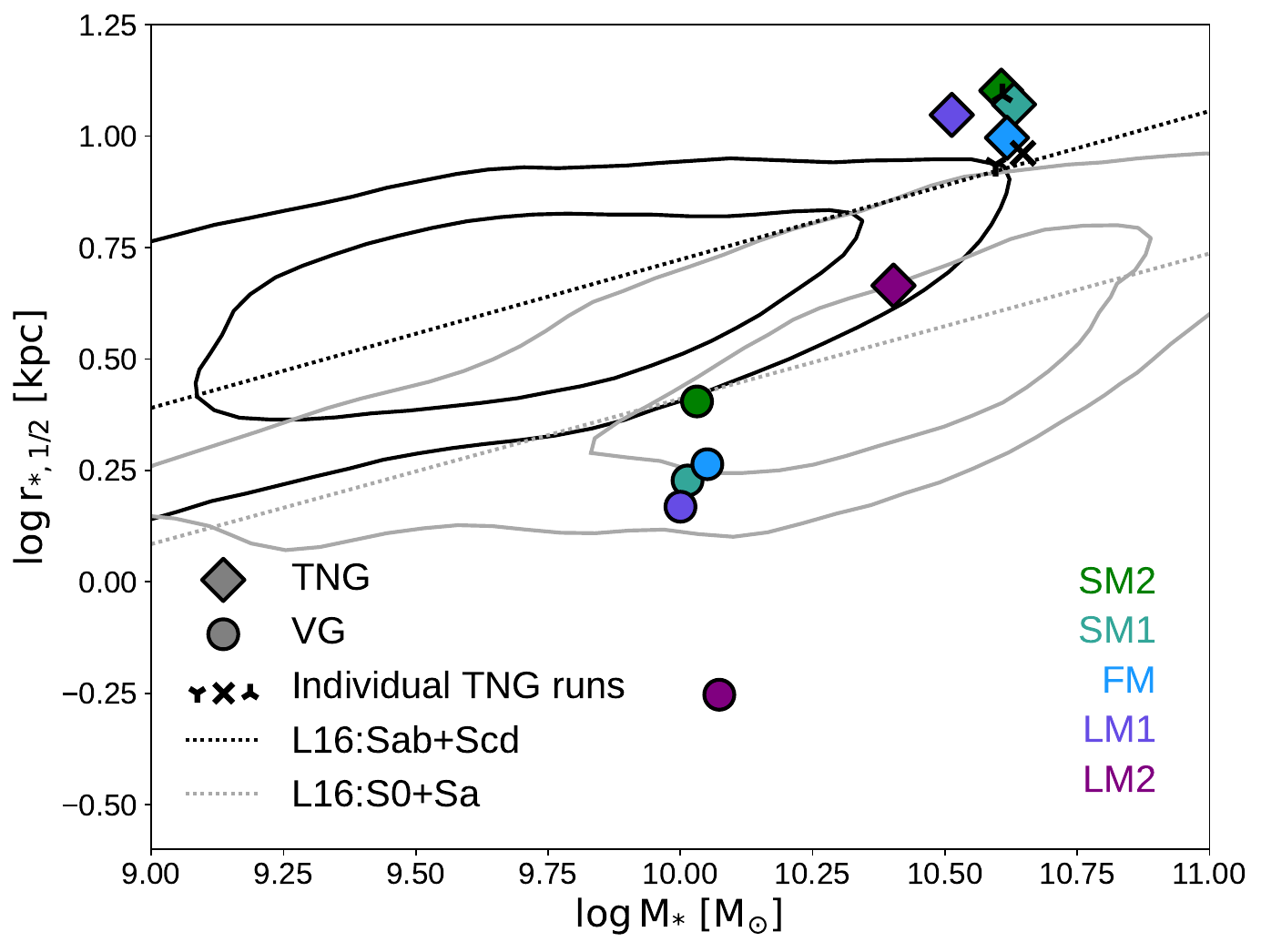}
    \caption{Location of the simulated galaxies on the size-mass plane, where size is measured as the \emph{R}-band half-light radius, while stellar mass is measured within $3 \times \RHALF$. The black and grey contours show the distribution of observed galaxies from the GAMA survey \citep[][L16]{Lange2016}, for their sample of Sab+Scd and S0+Sa galaxies respectively (the contours indicate the 50$^{\rm th}$ and 90$^{\rm th}$ percentiles of each distribution); the corresponding dotted lines show their best-fitting linear relations. The \vgn{} galaxy sizes are in agreement with the S0+Sa sample, albeit smaller than average. This reflects their proportionally larger bulges than the \tng{} galaxies, whose sizes are in agreement, although above average, with the Sab+Scd sample. The \vgn{} \XCXX{} is especially small and in fact beyond the distribution of even elliptical galaxies (not shown) in the L16 sample.}
    \label{fig:sizeMassRelation}
\end{figure}

\begin{figure*}
    \centering
    \includegraphics[width=\linewidth]{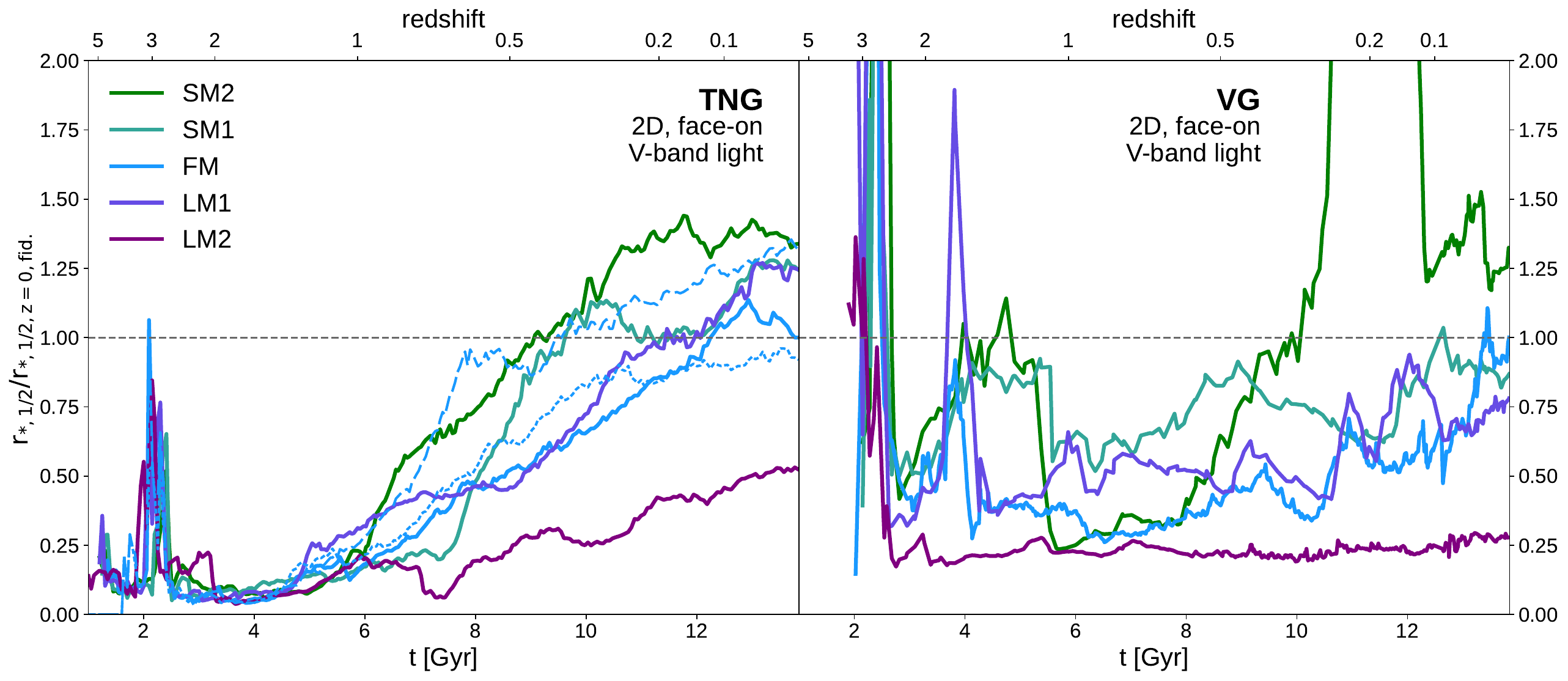}
    \includegraphics[width=\linewidth]{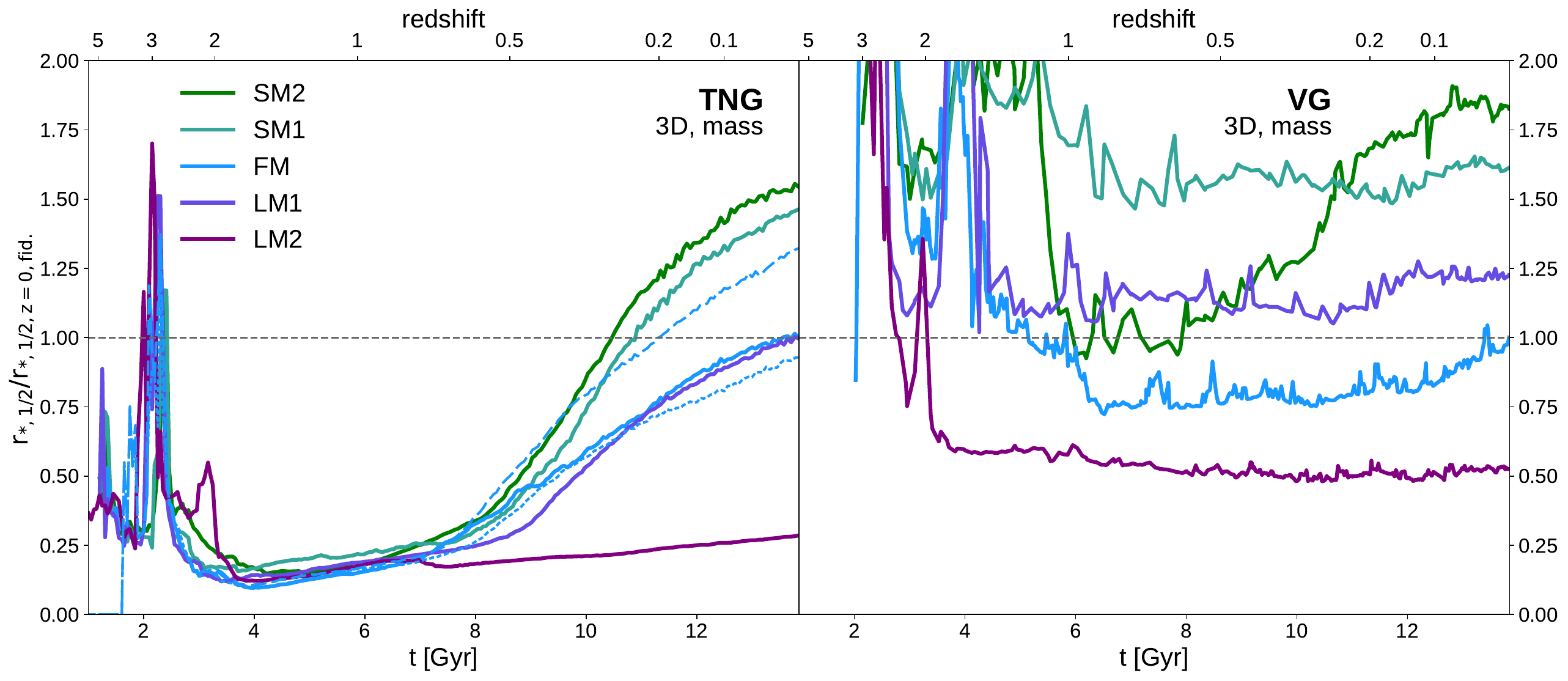}
    \caption{Evolution of 2D half-light (top) and 3D half-mass (bottom) radii for the \tng{} (left) and \vgn{} (right) sets of simulations. Each radius has been normalised by the $z=0$ radius of the fiducial run from the corresponding set. Bottom panels show the evolution of stellar mass within $3\,\RHALF$. The dotted and dashed blue lines in the left panels show the re-simulations of the fiducial ICs with the \tng{} model to show stochastic scatter (\XCRII{} and \XCRIII{} respectively). The \tng{} galaxies show a clear trend of more rapid size growth as a function of time for the smaller merger scenarios, for both half-light and half-mass radii, beginning at $z \sim 1$. Such a trend is also seen for the \vgn{} simulations when considering half-light radii; a trend is not obvious for the half-mass radii however, likely due to their larger bulge fractions (see Section \ref{sec:kinComp}). The rapid rise in half-light radius for the \vgn{} \XXC{} simulation at $z \sim 0.2$ is caused by an influx of gas resulting in SF in the outskirts of the galaxy. We have restricted the y-axis values to show the majority of the trends in all simulations more clearly and therefore do not show the peak of this feature, which is approximately at $\RHALF/\RHALFGEN{\rm z=0,fid.} = 5.9$ and $z=0.2$.} \label{fig:sizeEvolution}
\end{figure*}

\begin{figure*}
    \centering
    \includegraphics[width=\linewidth]{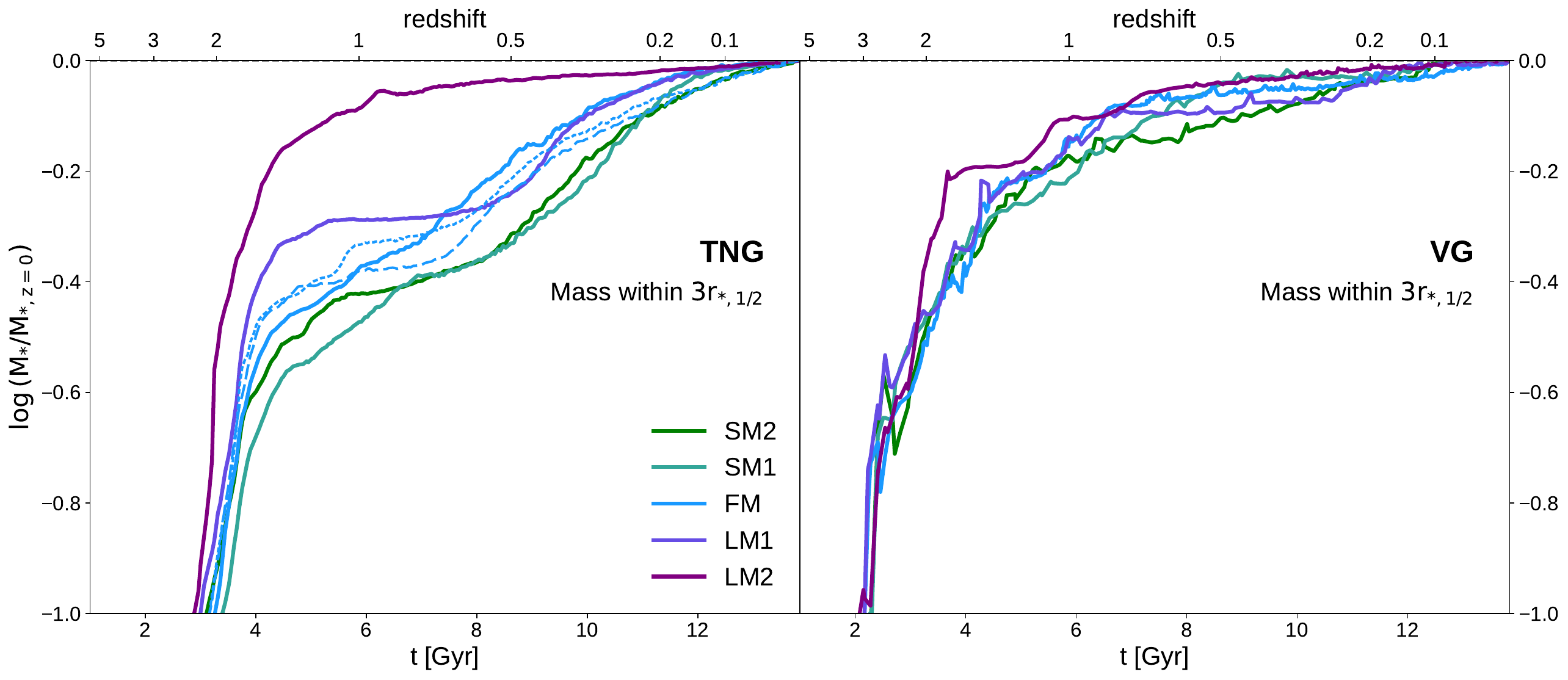}
    \includegraphics[width=\linewidth]{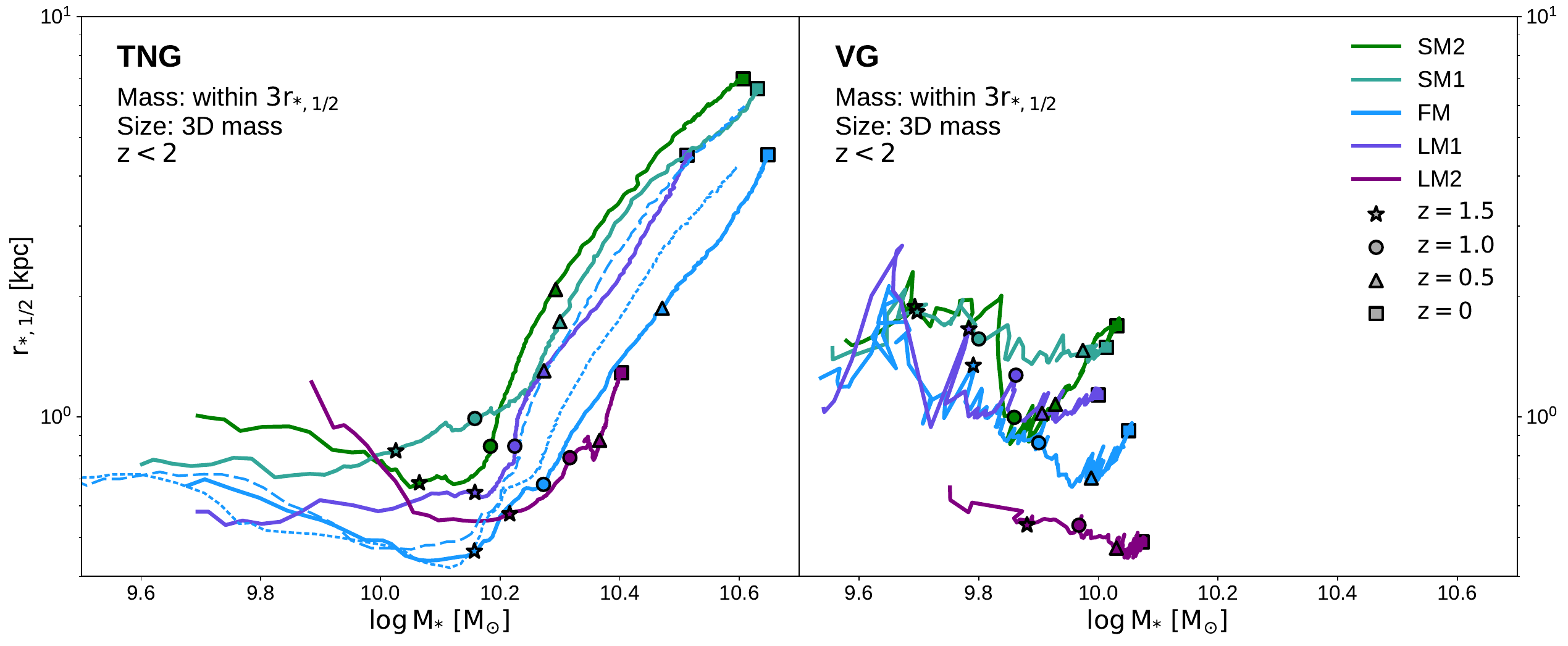}
    \caption{\emph{Top panels:} Evolution of galaxy stellar mass (measured within $3\,\RHALF$). Each of the curves has been normalised by the mass of the same galaxy (not the fiducial galaxy) at $z=0$. The dotted and dashed blue lines in the left panels show the \XCRII{} and \XCRII{} simulations as in Fig. \ref{fig:sizeEvolution}. \emph{Bottom panels:} Evolution of the galaxies in the size-mass plane. The results at $z>2$ are omitted for clarity. The various markers indicate the specified redshifts. There is a clear trend of smaller \targetMerger{} merger ratio resulting in larger galaxy sizes even at constant stellar mass in both sets of simulations, though differing significantly in absolute terms. The evolution in the size-mass plane shows two separate phases: a mild contraction phase where sizes decrease with increasing stellar mass (until $z \sim 1.5$ for the \tng{} simulations, $z \sim 0.5$ for the \vgn{} simulations), followed by an expansion phase where sizes increase with increasing mass. The \tng{} simulations show considerable growth in both mass and size in the expansion phase, whereas in the \vgn{} case, the degree of growth is similar to the degree of the previous contraction.} \label{fig:massEvolution}
\end{figure*}

We now turn to correlations between galaxy properties and merger histories to understand the response of the two codes to the genetic modifications in the ICs. The main aim of the modifications was to alter the mass ratio of the \targetMerger{} merger while attaining the same halo mass at $z=0$, in practice by altering the overdensity of the region at $z=99$. It is impossible to avoid changing other aspects of the merger however, as seen in Table \ref{tab:gmProperties}. Apart from the merger timings being mildly altered by up to a few 100 Myr, the orientation of the mergers is also affected. In each of the simulations, the \targetMerger{} merger is largely radial, as indicated by the $v_{\rm r}/v_{\theta}$ ratios at the time of accretion. However, differences in this ratio and the magnitude of the relative velocity at this time translate to significant differences in the impact parameters of the merger, with the FM galaxy having the most radially-oriented merger in both sets of simulations. This is an important factor to keep in mind when interpreting subsequent results.

\subsection{Galaxy sizes at $z=0$}

As seen from Fig. \ref{fig:stellarMocks}, the galaxies become smaller with increasing $z=2$ merger mass ratio. Therefore, we first examine the $z=0$ sizes of the galaxies, characterised by the stellar half-light and half-mass radii in Fig. \ref{fig:sizesToday}, normalised by the fiducial values. In the case of the fiducial \tng{} simulation, we show the average values for all three runs as the large blue diamond and normalise by this averaged value instead; the individual runs are shown with smaller cross, up and down markers. Both panels confirm that the galaxies become smaller when the $z=2$ merger has a larger mass ratio, albeit with some scatter such that the trends are not monotonic, particularly due to the intermediate merger scenarios of \XCX{} and \XVC{}. The trends are similar between the two simulation codes and marginally stronger when considering the total mass compared to V-band light. Furthermore, the deviations due to the altered merger histories are larger than the stochastic scatter (as measured with the \XCRII{} and \XCRIII{} runs, which is of order $\sim10\%$).

Fig. \ref{fig:sizeMassRelation} shows the location of the simulated galaxies in a size-mass plane. Here we show mass within $3 \times \RHALF$, while size is measured as the \emph{R}-band 2D half-light radius, measured in the same way as the \emph{V}-band half-light radius. We use the \emph{R}-band here in order to compare our results with observations (we have confirmed these are nearly identical to \emph{r}$_{\rm SDSS}$-band sizes for our sample). The contours show the distribution of observed galaxies from the GAMA survey \citep[][L16 hereafter]{Lange2016}, with the black and grey dotted contours representing their Sab+Scd and S0+Sa subsamples. The corresponding solid lines show their best-fitting linear relations. The \vgn{} galaxies are seen to be in good agreement with the L16 S0+Sa sample, albeit smaller than average. The \tng{} galaxies, being more massive, lie outside the 90$^{\rm th}$ percentile contours of the L16 distributions; however, this is due to the relatively small number of massive galaxies in the observations themselves, and the \tng{} galaxy sizes are in fact only slightly larger than the average massive Sab+Scd galaxies reported in L16. As we show in later sections, these trends are consistent with the fact that the \tng{} galaxies have a larger disc proportion (more akin to Sab+Scd type morphologies), compared to the \vgn{} galaxies which have more significant bulge contributions (thus more like S0+Sa type morphologies). The \XCXX{} galaxies in both sets of simulations are seen to be smaller than average, which is also consistent with them being bulge-dominated. In fact, the \vgn{} \XCXX{} galaxy lies outside the distribution of even elliptical galaxies (not shown) in the L16 sample, which are smaller than the disc samples reproduced here. In addition to morphology, the scatter in the the size-mass relation has also been shown to be impacted by ages and SFHs; \citet{Scott2017} show that high stellar surface brightness galaxies are older and more metal rich and alpha-enhanced. As shown in Fig. \ref{fig:fiducialSFR}, the \tng{} fiducial galaxy is indeed younger than the \vgn{} galaxy and we have confirmed that this is the case for all galaxies in our sample, consistent with the observed findings.

These results indicate that modifications to the \targetMerger{} merger are correlated with galaxy properties nearly 10 Gyr later, but do not establish a causal connection between the two. As discussed in previous work with these simulations, the modifications to a single merger necessitate other modifications to the overall merger and accretion history of the halo in order to produce the same halo mass at $z=0$. The late-time properties of the galaxies are impacted by this entire modified history \citep{Rey2023,Joshi2024}. 

\subsection{Evolution of galaxy sizes}

To understand how the modifications targeting a \targetMerger{} merger impact on present-day galaxy sizes, we next show the evolution of the galaxy sizes with cosmic time in Fig. \ref{fig:sizeEvolution}. The upper and lower panels show the half-light and half-mass radius respectively. The peak radii seen at $z=3$ in both sets of simulations are likely to be significantly overestimated, since the galaxies undergo a major merger at this time, preceding the \targetMerger{} merger. In the \tng{} case, differences in galaxy sizes set in at $z \sim 1.3$ for $\RHALFV$ and $z \sim 0.7$ for $\RHALF$. The galaxies continue to grow at later times, with the growth rate also being higher for a smaller \targetMerger{} merger mass ratio, resulting in the $z=0$ trends. The time at which this accelerated growth is seen in the \tng{} simulations coincides approximately with the beginning of the period where the \tng{} fiducial galaxy consistently has a nearly 1 dex higher SFR compared to the \vgn{} galaxy. This enhanced SF therefore appears to be responsible for the rapid growth in size for the \tng{} galaxies. The larger fluctuations in the half-light radius compared to the half-mass radius are likely due to the former being highly sensitive to the presence of non-uniformly distributed young stars.

The \vgn{} simulations show similar trends with the half-light radii growing steadily after $z\sim1$. However, there are some significant differences: Unlike in the \tng{} vase, the \vgn{} galaxies establish significant size differences around $z=2$, and there is little growth in the half-mass radius after $z\sim 1.5$, except in the \XXC{} case. An important factor to consider here is that, as we show in the following sections, the \vgn{} galaxies have a higher proportion of their mass in the bulge compared to the \tng{} galaxies (while still being disc-dominated overall) and a markedly different mass profile. Therefore the $\RHALF$ measurement has a different contribution from the bulge between the two codes. In Appendix \ref{sec:appendixSizes}, we discuss the impact of measuring galaxy sizes with $\RHALF$ and the impact of the two components, but summarise here that we do not see a qualitative difference when using disc sizes alone.

The \vgn{} \XXC{} simulation also shows a feature around $z\sim0.2$ where it has a significantly larger half-light radius for a nearly 2~Gyr period. We have confirmed that this is due to SF occurring in the outskirts of this region following an influx of gas, which does not greatly alter the mass distribution of the galaxy, but does impact the light distribution, particularly in the V-band. Aside from this exception, we find that the overall trends seen at $z=0$ were established at $z\sim1$ even in the \vgn{} simulations. 

To confirm that the trends in size evolution are not simply due to trends in stellar mass evolution, we show the galaxies' stellar mass as a function of time, normalised by their own mass at $z=0$ in the upper row of Fig. \ref{fig:massEvolution}. The galaxies with the larger merger scenarios build up their stellar mass earlier in both suites, although the trend is more pronounced in the \tng{} case. In the lower panels of Fig. \ref{fig:massEvolution}, we show the loci of the galaxies in the size-mass plane. Here we show absolute sizes and masses to make it easier to compare the scaling relations to each other. We omit data points at $z>2$ to reduce noise stemming from the early history of the galaxies when they are undergoing multiple mergers. The figures confirm that the correlation between sizes and \targetMerger{} merger ratios are not simply due to an underlying correlation with stellar mass; the smaller merger scenarios result in larger galaxies at fixed mass. Although the two sets of simulations appear to have significantly different evolution in this plane, one can discern two distinct phases -- a `contraction' phase where the sizes become smaller with increasing stellar mass, and an `expansion' phase where the sizes increase with increasing stellar mass. In the \tng{} simulations the `expansion' phase begins at $z \sim 1.5$ (star markers) characterised by rapid and significant growth, such that for all but the \XCXX{} case, a 0.5 dex growth in stellar mass corresponds to a factor of 5-7 increase in half-mass radius. On the other hand, in the \vgn{} simulations, the `expansion' phase is seen to begin only at $z\sim 0.5$ (triangle markers) and have a $\sim 0.05-0.1$ dex growth in stellar mass correspond to at most a factor of $\sim 1.8$ (and usually less) increase in half-mass radius. Interestingly, during the `contraction' phase, the sizes from both sets of simulations are much more comparable; in fact, the fiducial \tng{} galaxy is smaller than the fiducial \vgn{} galaxy at a given stellar mass. 

In summary, despite the different numerical methods and galaxy formation models involved, the two simulation codes have similar responses to the modified ICs, with the consequence that increasing the merger mass ratio of a $z=2$ merger leads to a larger galaxy nearly 10 Gyr later at $z=0$ due to the overall modifications made to the accretion history of the galaxy. The particular way in which the sizes are established however, differs between the \tng{} and \vgn{} simulations -- while the former grows markedly after $z=2$, the latter has a more modest growth -- necessitating further investigation to understand the origin of the size trends in each case. In the next section, we examine the kinematics of the galaxies to understand how the galaxies' sizes grow and what commonalities exist between the two sets of simulations.

\section{Size growth through disc growth} \label{sec:kinComp}

\begin{figure*}
    \centering
    \includegraphics[width=\linewidth]{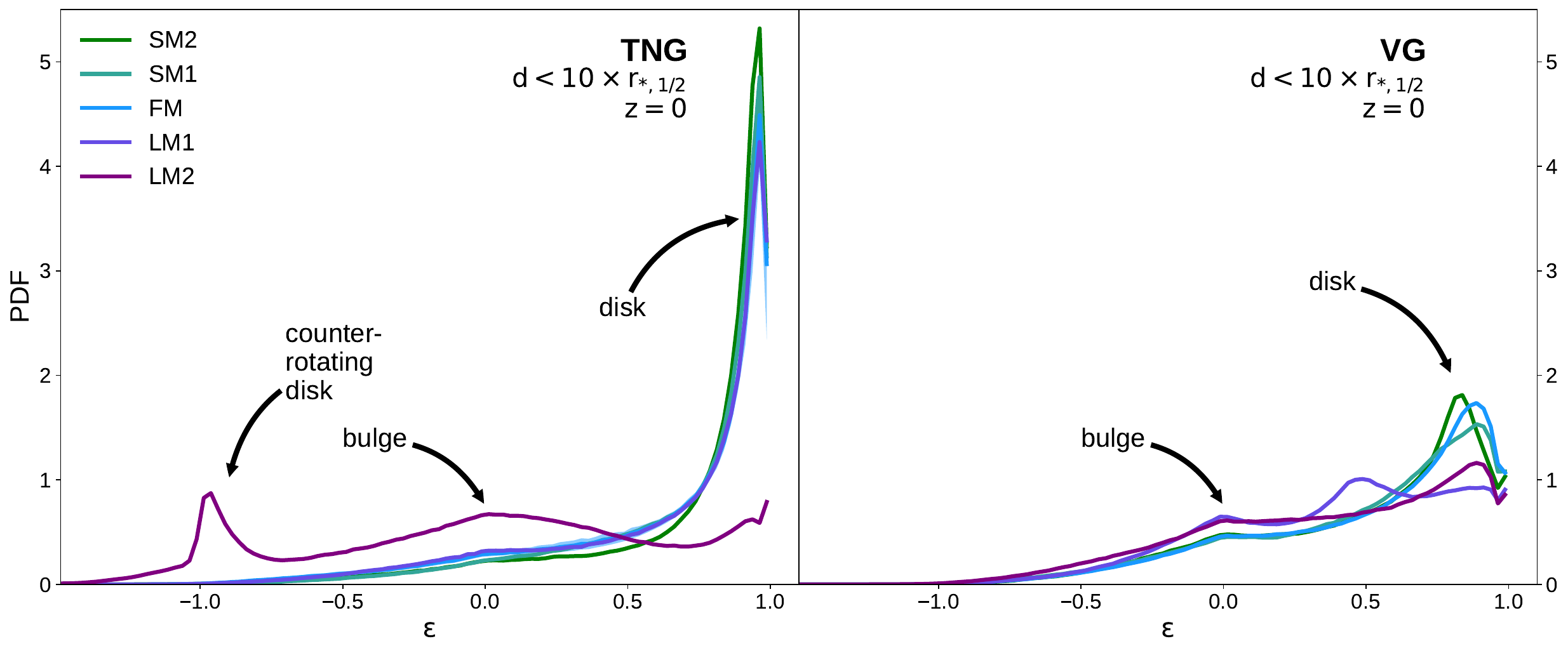}
    \caption{Mass-weighted probability distribution functions of the circularity parameter $\epsilon$ for all stellar particles within $10\,\RHALF$ at $z=0$. Both sets of simulations show a prominent disc component at $\epsilon \approx 0.9$ and a secondary bulge component at $\epsilon=0$, except for the \XCXX{} cases which are bulge-dominated galaxies. Furthermore, the smaller merger scenarios result in a relatively more massive disc component. In the \tng{} simulations, the disc component is proportionally more important and corresponds to a kinematically colder disc than in the \vgn{} simulations (the peak of the distribution corresponding to the disc is at a slightly higher value of $\epsilon$ for the \tng{} simulations). Two other important features are evident in the distributions: a `counter-rotating disc' in the \tng{} \XCXX{} simulation, and a `disturbed disc' in the \vgn{} \XCX{} simulation.}
    \label{fig:circDistributionZ0}
\end{figure*}

\begin{figure*}
    \centering
    \includegraphics[width=\linewidth]{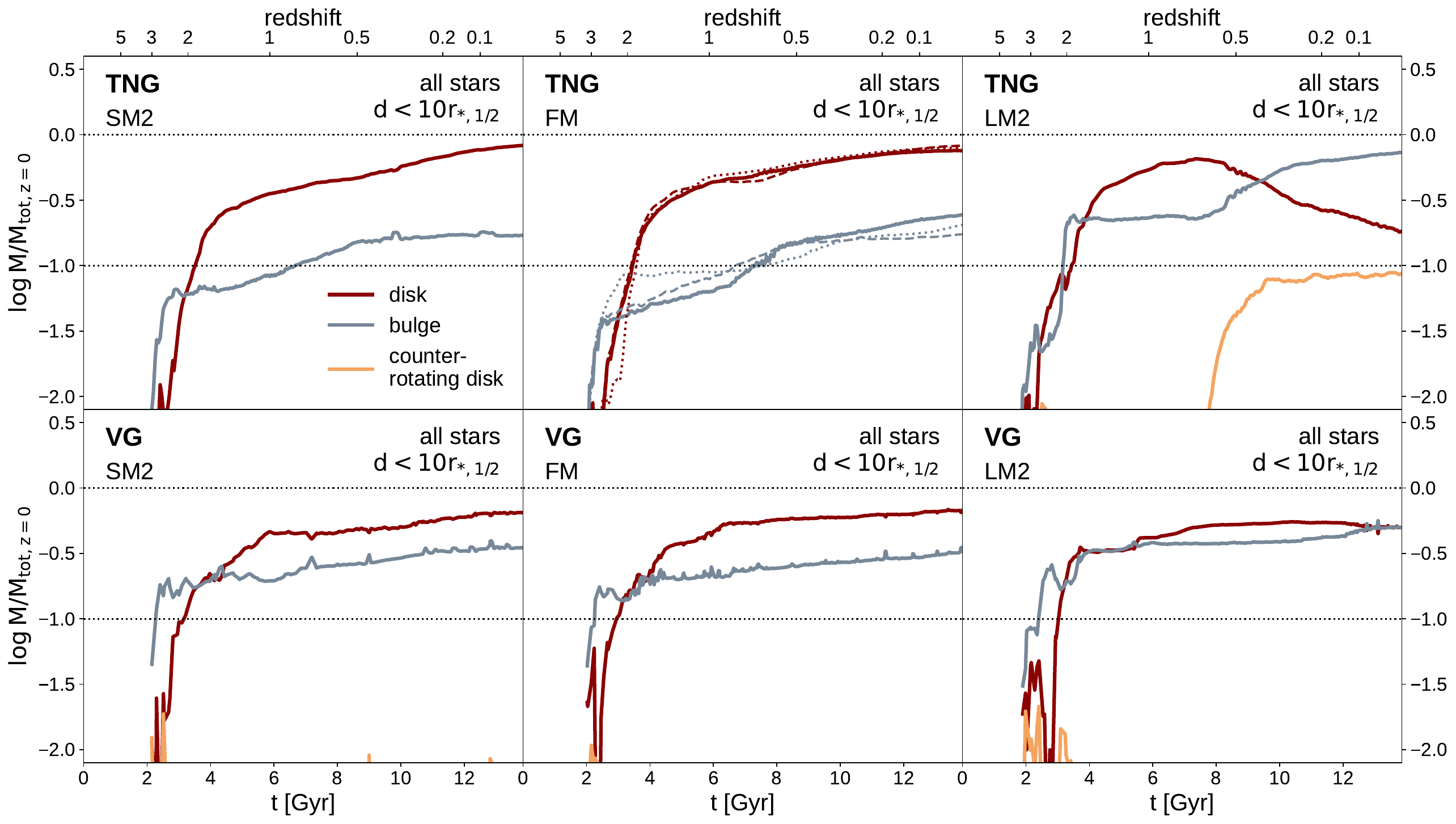}
    \caption{Evolution of the stellar mass in disc (red), bulge (gray) and counter-rotating disc (orange) components within $10\,\RHALF$ as a function of cosmic time, for the \tng{} (top row) and \vgn{} (bottom row) simulations. The masses have been normalised by the galaxies' $z=0$ stellar mass within $10\,\RHALF$. We omit the two intermediate runs for clarity. The two additional runs of the \tng{} fiducial simulations (\XCRII{} and \XCRIII{}) are shown with dotted and dashed curves in the top middle panel. Dotted black lines indicate 10 and 100 percent of $M_{\rm tot,z=0}$ for reference. In the \tng{} simulations, the increase in \targetMerger{} merger mass ratio results in a proportionally higher build-up of bulge mass compared to disc mass. In the \vgn{} simulations, although the trends are not as clear. Additionally in the \tng {} \XCXX{} case, we also see the rapid build-up of the counter-rotating disc component at $z \sim 0.5$, which then accounts for nearly 10 per cent of the galaxy mass. The simultaneous decrease in disc mass is not accompanied by a loss of total stellar mass, but rather due to the transfer of disc stars to the bulge component. In the \vgn{} \XCXX{} case, we find a similar relative increase in bulge mass, resulting in an equal proportion of disc and bulge mass at $z=0$.} \label{fig:circCompEvolution_allStars}
\end{figure*}

\begin{figure*}
    \centering
    \includegraphics[width=\linewidth]{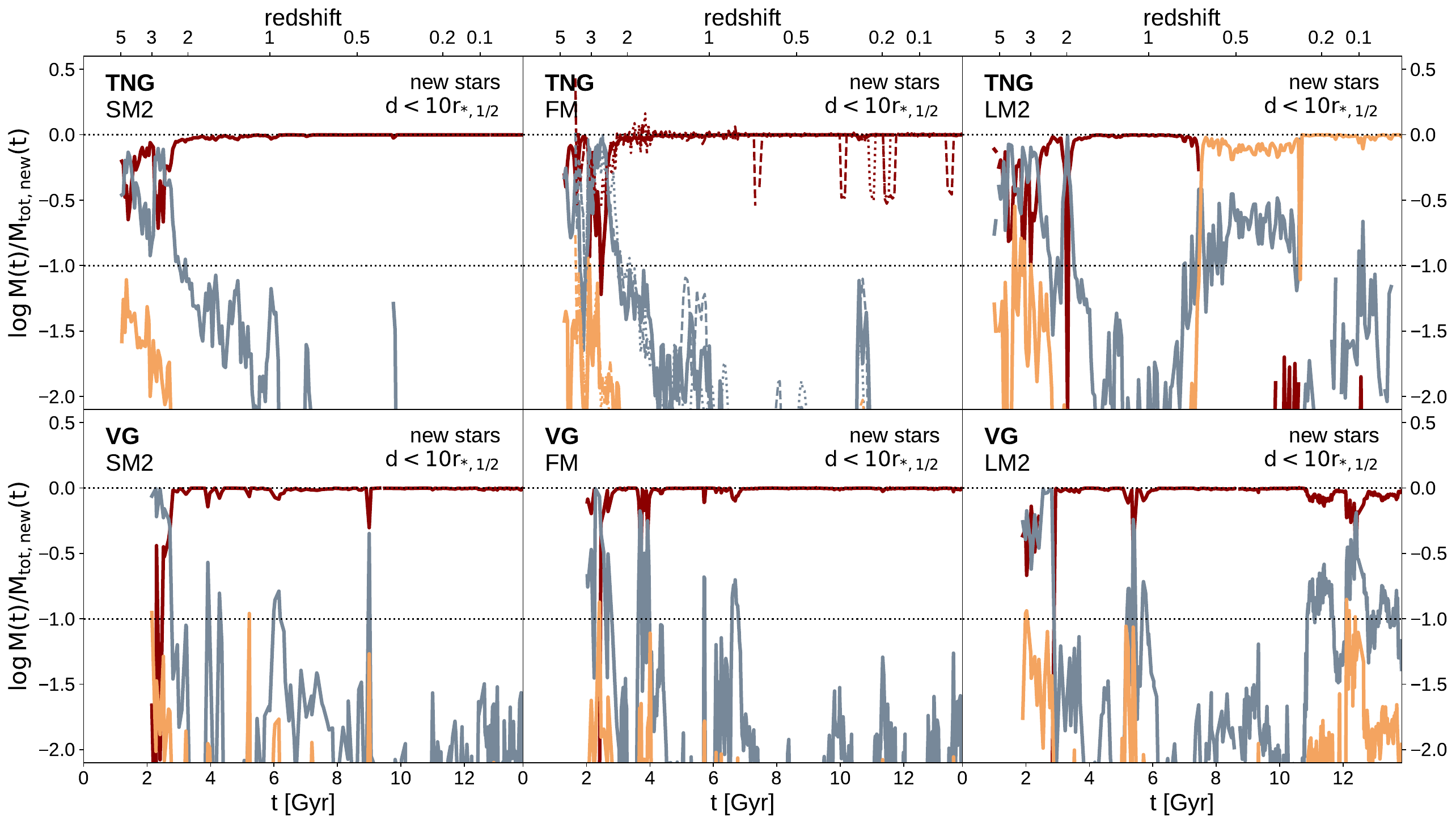}
    \caption{Evolution of the stellar mass in disc (red), bulge (gray) and counter-rotating disc (orange) components as in Fig. \ref{fig:circCompEvolution_allStars}, but here considering only new stars (formed within the last 50~Myr) as a proxy for in-situ SF. The masses are shown as a fraction of the total mass in new stars the time, to emphasize the relative importance of each component in ongoing SF. Dotted black lines indicate 10 and 100 percent of $M_{\rm tot,new}(t)$ for reference. In the fiducial and smaller merger scenarios, the disc component at each snapshot accounts for nearly all the in-situ SF after $z \sim 2$ in both sets of simulations. In the \tng{} \XCXX{} case however, after $z \approx 0.7$, nearly all the in-situ SF occurs in the counter-rotating disc component, although the bulge component can account for up to 10 per cent by mass. Similarly, in the \vgn{} \XCXX{} simulation, we find a considerable increase in the contribution to in-situ SF from the bulge component starting at $z\sim 0.25$, accounting for 10-30 per cent, although the disc component still remains dominant.} \label{fig:circCompEvolution_newStars}
\end{figure*}

\begin{figure*}
    \centering
    \includegraphics[width=\linewidth]{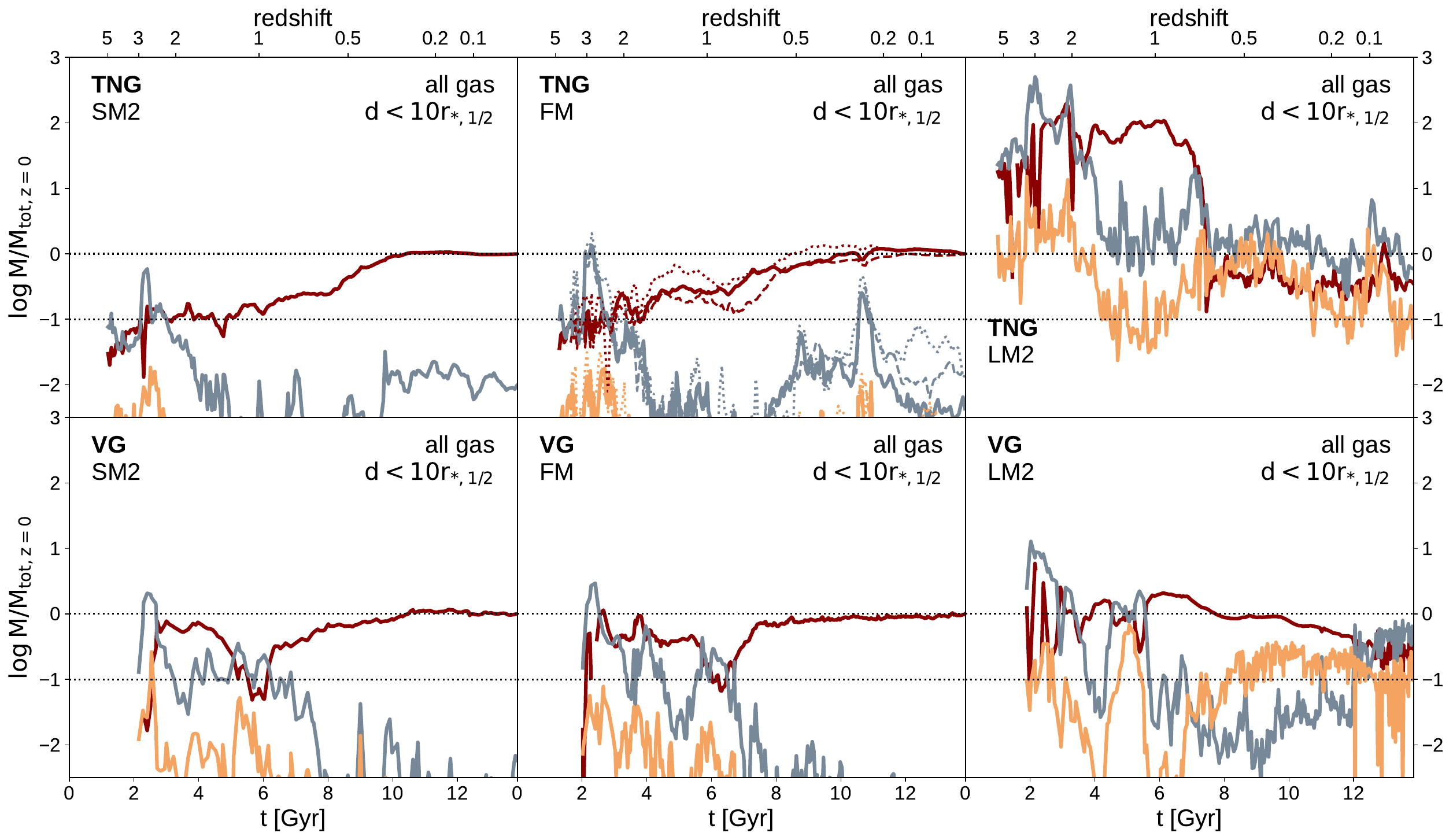}
    \caption{Evolution of the gas mass in disc (red), bulge (gray) and counter-rotating disc (orange) components as in Fig. \ref{fig:circCompEvolution_allStars}. The masses have been normalised by the galaxies' $z=0$ gas mass within $10\,\RHALF$. Dotted black lines indicate 10 and 100 percent of $M_{\rm tot,z=0}$ for reference. In the fiducial and smaller merger scenarios, the majority of gas settles into a disc configuration at early epochs -- by $z \sim 2$ in the \tng{} simulations, $z \sim 1$ in the \vgn{} simulations -- and fuels the SF seen in Fig. \ref{fig:circCompEvolution_newStars}. In the \tng{} \XCXX{} simulation, we see a rapid re-arrangement of the gas at $z \sim 0.7$ such that the disc component becomes subdominant to the bulge and counter-rotating components. Similarly, in the \vgn{} \XCXX{} simulation, although the disc component remains the dominant one, there is a significant increase in the counter-rotating disc mass starting at $z \sim 0.7$ and bulge mass in the last 2~Gyr. These latter scenarios are responsible for the \XCXX{} galaxies being bulge-dominated in both sets of simulations.} \label{fig:circCompEvolution_gas}
\end{figure*}

\begin{figure*}
    \centering
    \includegraphics[width=\linewidth]{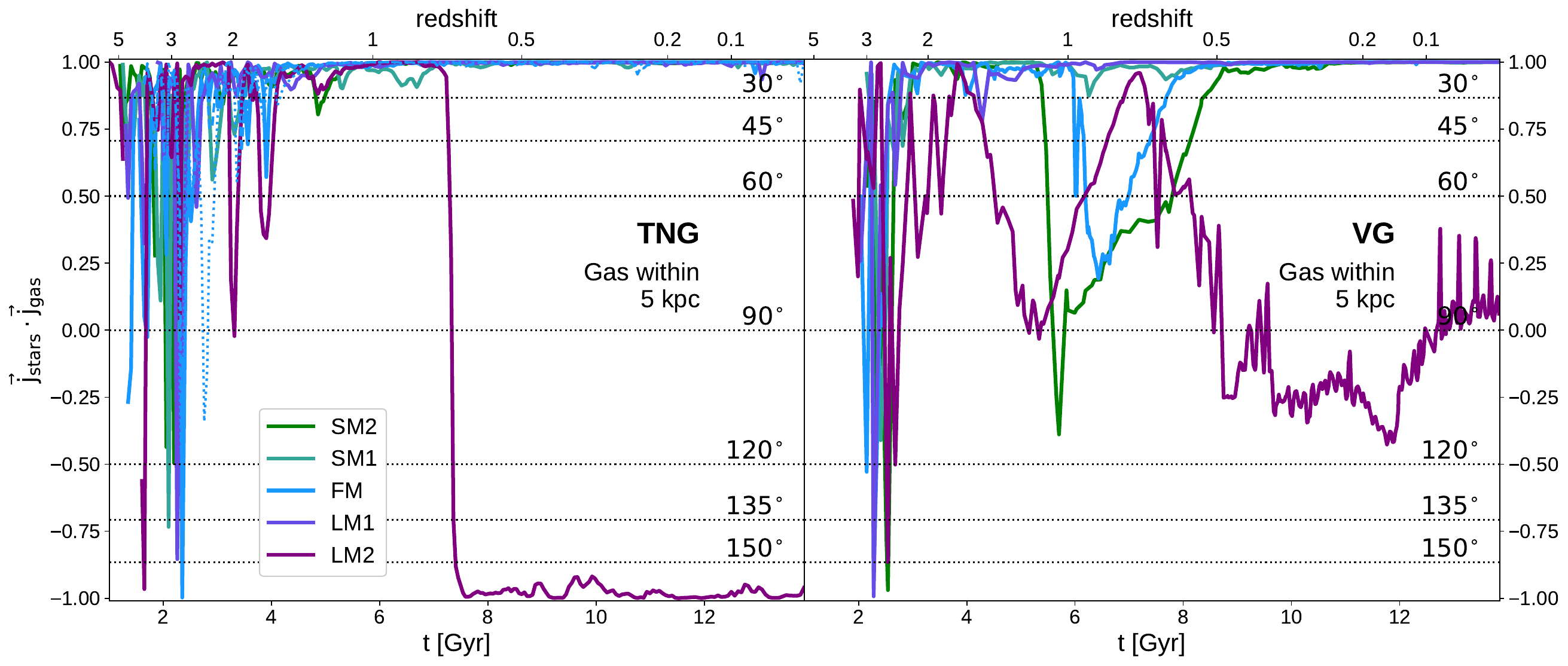}
    \includegraphics[width=\linewidth]{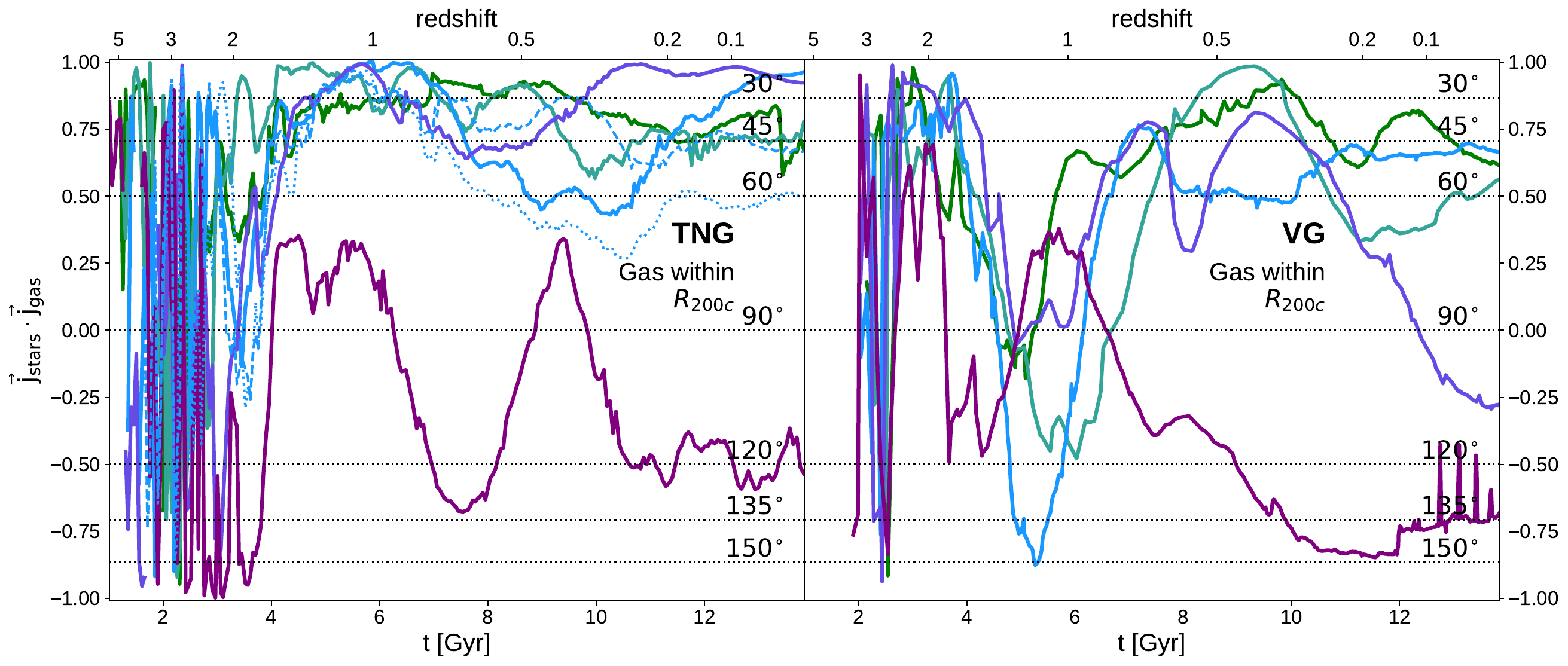}
    \caption{The alignment between the gas and stellar angular momentum as a function of time for the two sets of simulations, measured as the dot product between the respective unit angular momentum vectors. The stellar angular momentum is measured using particles within 5~kpc. Gas angular momentum is measured within 5~kpc (top row) and within $\RHOST$ (bottom row). The dotted and dashed blue curves in the left panels show the results for the \tng{} \XCRII{} and \XCRIII{} simulations. Black dotted horizontal lines indicate various angles between the two vectors to aid comparisons. In the \tng{} simulations, the gas angular momentum in the inner regions is well-aligned to the stellar angular momentum, with the exception of the \XCXX{} case, where there is an influx of perfectly counter-rotating gas at $z \approx 0.8$. In the \vgn{} simulations, we see the same influx of misaligned gas, although here, its angular momentum is approximately perpendicular to the stellar angular momentum rather than being perfectly anti-aligned. We also see the presence of misaligned gas at earlier times, except for the \XVC{} and \XCX{} cases, which may account for the overall lower disc fractions of the \vgn{} simulations.} \label{fig:gasStarAlignment}
\end{figure*}

\subsection{Kinematic decomposition at $z=0$} \label{sec:kinDecompZ0}
As can be seen in Fig. \ref{fig:stellarMocks}, the four merger scenarios apart from \XCXX{} result in a disc-dominated central galaxy at $z=0$ in both sets of simulations, while \XCXX{} results in a bulge-dominated galaxy (albeit with a significant disc component in the \tng{} simulations). It is therefore instructive to explore the connection between galaxy size and the prominence of the disc component, as a means of understanding how both simulations grow in response to the increased/decreased $z=2$ merger mass ratio. 

We use the circularity parameter $\epsilon$ (see Section \ref{sec:postprocessing}) to characterize the kinematics of the galaxies. Fig. \ref{fig:circDistributionZ0} shows the distribution of $\epsilon$ for all galaxies in our sample at $z=0$, including stellar particles within $10 \times \RHALF$. Particles from such large radii are included in these figures to highlight some important kinematic components that may impact the size trends in the previous section. We have confirmed that the results within $3\,\RHALF$ are qualitatively the same in most cases and we highlight any key differences below. 

In both sets of simulations, two major components are seen in Fig. \ref{fig:circDistributionZ0} -- a narrower peak at high values of $\epsilon$ corresponding to the `disc' component, and a broader peak at $\epsilon \sim 0$ corresponding to the `bulge' component. Significant differences can be seen between the two sets of simulations -- the \tng{} galaxies are more disc-dominated and have kinematically colder discs compared to the \vgn{} galaxies (as indicated by the marginally higher value for the peak of the $\epsilon$ distribution corresponding to the disc component), with the exception of the \XCXX{} galaxies. This may be an important contributing factor in the \tng{} galaxy sizes being significantly larger compared to the \vgn{} galaxies, as discussed in detail in Section \ref{sec:discModelDiffs}. Despite these overarching differences, there are also similarities in the response of the simulations to the modified ICs. The increase in \targetMerger{} merger mass ratio results in an increase in the mass fraction of the bulge component through the inevitable alterations to the overall accretion history of the galaxies. In fact, in the \XCXX{} case, the bulge component accounts for 50 and 73 per cent of the stellar mass within $10\,\RHALF$ for the \vgn{} and \tng{} galaxies respectively. In the other four cases, the disc component dominates, with the disc proportion being higher for the \tng{} simulations (74-84 per cent) than the \vgn{} simulations (57-67 per cent). 

Apart from the two main components, i.e. the disc and bulge, there are two other features highlighted in Fig. \ref{fig:circDistributionZ0}. The first is seen for the \XCXX{} case in the \tng{} simulations, where there is another significant peak in the distribution at $\epsilon \sim -1.0$, which corresponds to a `counter-rotating disc' component. Note that this peak is significantly smaller when considering stellar particles only within $3\,\RHALF$, indicating that this component is mostly found in the outskirts of the galaxy. We discuss its origins in more detail in the following sections. The second key feature is seen in the \XCX{} case in the \vgn{} simulations, where the peak at $\epsilon \sim 0.9$ is significantly smaller than the other simulations, but accompanied by a peak at $\epsilon \sim 0.5$. This latter peak is due to the influx of misaligned gas and subsequent SF with the end result being that a previously discy component becomes more dispersion dominated, and the peak of the distribution moves from $\epsilon \sim 0.9$ to 0.5. We have confirmed that the feature is long-lived and can be traced back to approximately 3~Gyr ago. We therefore refer to this component as a `disturbed disc'. This disturbed disc component is prominent when only considering particles within $3\,\RHALF$, indicating that the disruption to the disc is not restricted to the outskirts of the galaxy. Hence, the disruption is due to sustained SF from misaligned gas within the main body of the galaxy rather than an influx of ex-situ stellar particles with misaligned orbits.

\subsubsection{The impact of AGN feedback in the \tng{} simulations} \label{sec:agn}
The presence of AGN feedback in the \tng{} model plays an important role in establishing the kinematic make-up of the galaxies and their sizes \citep[e.g.][]{TNGMethodsPillepich2018,Habouzit2019}. As we discuss in later sections, the modifications to the ICs also impact the accretion of gas into the galaxy, likely affecting the gas accretion onto the central SMBH and the resultant AGN feedback. A detailed analysis on this topic is beyond the scope of this paper, but here we present some preliminary findings. The GMs are indeed seen to impact BH growth in the \tng{} galaxies, with the largest and smallest merger scenarios resulting in the most and least massive BHs, not only at $z=0$ but at all times since $z \sim 2$. The \XC{} galaxy has a BH mass of $7.0\times 10^{7}\MSUN$, while the \XXC{} and \XCXX{} galaxies have BHs of mass $5.7\times 10^{7}\MSUN$ and $8.0\times 10^{7}\MSUN$ respectively. There is significant scatter in this relation however, as evidenced by the two intermediate GM simulations as well as the two additional runs of the fiducial ICs. It has been shown in \citet{Davies2022,Davies2024}, with the use of the GM technique in the EAGLE model \citep{Crain2015,Schaye2015}, that major mergers are likely to boost the growth of a central BH by disrupting the stellar disc and thus the gaseous disc surrounding the halo, whereas less disruptive mergers do not affect the growth of the BH. Our results support this scenario, with the largest \targetMerger{} scenario leading to the most massive BH. 

Several studies have shown that the absence of AGN feedback results in massive galaxies ($\MSTAR \gtrsim 10^{10}\MSUN$) being too compact \citep[e.g. see][]{TNGMethodsPillepich2018,Cochrane2023}. The differences between our \vgn{} and \tng{} results is in agreement with this, although the presence/absence of AGN is one of several changes and therefore this particular comparison should be interpreted with care. \citet{Grand2017} find that the scale-length of the disc component is larger for their simulation without AGN feedback, but that it also has a more prominent bulge component, and thus the net effect of AGN feedback is a balance between these two somewhat competing trends. Indeed, the more massive BHs in our sample are correlated with \emph{smaller} disc sizes, similar to the anti-correlation found by \citet{Grand2017} between the degree of AGN feedback and galaxy size. This suggests that while the presence of any AGN feedback is important for producing larger galaxies, the impact of AGN feedback on disc sizes is modulated by other factors, some of which we discuss in Sections \ref{sec:kinDecompEvol} and \ref{sec:gasStarAlignment}. Furthermore, the stellar masses of our galaxies correspond to the precise mass at which the AGN feedback in the TNG model has been shown to transition from the thermal to kinetic mode, the latter of which is more efficient in driving outflows. Understanding the impact of the GMs on AGN feedback, and consequently on galaxy sizes, therefore requires careful consideration of the gas fuelling BH growth, the presence of a disc preventing gas accretion onto the BH, and the AGN feedback affecting SF in the galaxy and altering disc growth, which we will pursue in a future paper.

\subsection{Evolution of disc, bulge and counter-rotating components} \label{sec:kinDecompEvol}

In order to understand the connection between the galaxy sizes and their kinematics, we next investigate the evolution of the mass in the various kinematic components. Previous work including \citet[][]{Marinacci2014,Genel2015,Joshi2020} have defined the `disc' and `bulge' masses as the mass of particles with $\epsilon>0.7$ and twice the mass of particles with $\epsilon<0$. While these definitions are usually an adequate measure of the disc and bulge components, they do not effectively characterise the circularity parameter distributions for the galaxies in our samples, as demonstrated in Fig. \ref{fig:circDistributionZ0}. The presence of a `counter-rotating' disc leads to an over-estimation of the bulge component using the standard definition. Simultaneously, the disc component is seen to have a significant contribution even at $\epsilon<0.7$, especially in the \vgn{} simulations, and particularly in the presence of the `disturbed disc' component. For these reasons, we employ a modified measure of the mass in the disc, bulge and counter-rotating disc components, as follows:
\begin{itemize}
    \item Under the assumption that the `bulge' particles are distributed symmetrically around $\epsilon=0$, we first mirror the measured PDF between $-0.5<\epsilon<0$ around $\epsilon=0$ and fit a Laplace profile to this mirrored distribution between $-0.5<\epsilon<0.5$, thus excluding any disc or counter-rotating disc components (the Laplace profile was chosen empirically as it fit the data better than a normal distribution). 
    \item We initially measure a `bulge mass' as the integral of the predicted bulge PDF between $-1<\epsilon<1$.
    \item The `disc mass' and `counter-rotating disc mass' are then defined as the integral of the \emph{positive} difference between the measured PDF and the predicted bulge PDF between $0<\epsilon<1$ and $-1<\epsilon<1$ respectively.
    \item Finally, we obtain disc, bulge and counter-rotating disc fractions by normalising the corresponding masses by the sum of all three.
\end{itemize}
The mass of a given component is then defined as the fraction of the component multiplied by the total mass within the given aperture. We emphasise that this method is not designed to perfectly delineate the various kinematic components, but rather to provide a more accurate measure of their relative contributions than one obtained from absolute limits in $\epsilon$. Furthermore, Fig. \ref{fig:circDistributionZ0} shows that there is significant overlap between the bulge and disc components in the $\epsilon=0.2-0.5$ region, especially for the \vgn{} galaxies whose discs are characterized by a broader distribution in $\epsilon$. While our method aims at better separating the two components, it may over-estimate the disc mass in some cases, a factor we discuss in later results.

In Fig. \ref{fig:circCompEvolution_allStars}, we show the stellar mass of the three components as a function of time, including all stellar particles within $10\,\RHALF$. We omit the intermediate GM runs (\XVC{} and \XCX{}) for clarity and only show the fiducial and two extreme simulations. In both sets of simulations, in the fiducial and smaller merger scenarios, the disc mass is dominant at all times shortly after $z=2$. The increasing disc fractions with decreasing $z=2$ merger mass ratio are due to a combination of the growth of the disc mass especially at $z<0.5$ and a relatively lesser growth of the bulge mass; the trends in the \vgn{} simulations are weaker, as is the case in Fig. \ref{fig:circDistributionZ0}.

Furthermore, the counter-rotating disc component remains less than 1 per cent of the galaxy mass at all times for all but the \XCXX{} simulations. The \XCXX{} galaxy in the \tng{} simulations becomes bulge-dominated due to the rapid increase in the bulge mass and the simultaneous loss of disc mass beginning shortly before $z<0.5$, accompanied by a build up of the counter-rotating component which accounts for nearly 10 per cent of the mass of the galaxy within $10\,\RHALF$ (within $3\,\RHALF$, it only accounts for approximately 1 per cent of the mass). In the \vgn{} \XCXX{} simulation, although the magnitude of disc mass loss and bulge mass gain is more moderate, the trends are qualitatively the same, such that by $z=0$, the two contribute nearly equally to the total mass of the galaxy. However, we do not see a significant counter-rotating disc component as with the \tng{} simulation.

To determine whether these results are due to the accretion of ex-situ stellar material or due to in-situ SF, in Fig. \ref{fig:circCompEvolution_newStars}, we show the mass in the three kinematic components, but now considering only new stars, defined as those formed within the last 50~Myr as a proxy for in-situ SF. The masses are now normalised by the total mass of new stars at the given time, to highlight the importance of each of the kinematic components for in-situ SF. Additionally, in Fig. \ref{fig:circCompEvolution_gas}, we also show the evolution of the gas separated into the same kinematic components as for stars in Fig. \ref{fig:circCompEvolution_allStars}. Note that in measuring $\epsilon$ for the new stars and gas, we do not redefine the plane of the galaxy based on their angular momentum; the plane of the galaxy is always defined by the AM of all stellar particles within 5~kpc (see Section \ref{sec:postprocessing}).

In the fiducial and smaller merger scenarios, from Fig. \ref{fig:circCompEvolution_newStars} it is clear in both sets of simulations that in-situ SF occurs in the disc component as expected, while any SF in the bulge components is insignificant after $z \sim 2$. Similarly, in Fig. \ref{fig:circCompEvolution_gas}, gas within $10\,\RHALF$ is seen to settle into a mostly discy configuration after $z \sim 2$ in the \tng{} simulations, $z \sim 1$ in the \vgn{} simulations, for the fiducial and smaller merger cases. Therefore the discy in-situ SF results from gas in a discy configuration, as expected. However, the build-up of the stellar bulge at $z\sim 3$   (Fig. \ref{fig:circCompEvolution_allStars}) is only partially explained by star formation on bulge-like orbits. Since the galaxies undergo a major merger at $z \sim 3$, this mismatch indicates that the majority of the bulge component at this time is formed by the accreted stellar material brought in by the merger and the galaxies' own stellar material also being dispersion dominated.

The \XCXX{} case is markedly different in that in the \tng{} simulation, where the majority of in-situ SF after $z\sim 0.7$ occurs in the counter-rotating disc component as well as a significant amount in the bulge component, consistent with the presence of gas that is counter-rotating or dispersion-dominated beginning at $z=0.7$. Similarly, in the \vgn{} \XCXX{} simulation, even though the majority of in-situ SF occurs in the disc component, we do see a marked increase in SF in the bulge and counter-rotating disc components at $z<0.25$, accounting for 10-30 per cent of the mass in new stars, consistent with the gas in these components becoming dominant at this time. Note that the \tng{} \XCXX{} galaxy appears to lose a significant amount of gas mass at $z\sim 0.7$; this gas is not lost from the halo, but expelled to larger radii beyond the aperture under consideration. 

These results confirm that despite the significant differences between the two models, SF proceeds in a similar manner, with new stars inheriting the kinematic properties of the gas they are formed from, as expected. Importantly, they show that the kinematics of the gas within the galaxy follow a similar evolution in both sets of simulations. In addition to in-situ SF, the disc/bulge components may also be built up through the transfer of stellar particles across these components, although we have not quantified the contribution of the two channels in building the disc/bulge fractions. Preliminary analysis (not shown) indicates that even in the four smaller \targetMerger{} merger scenarios where the bulge component is always subdominant to the disc, there is a significant transfer of stellar particles from the disc to the bulge. We will explore the relative importance of these channels in a future paper.

\subsection{Alignment of stars and gas} \label{sec:gasStarAlignment}

To probe the connection between the kinematics of the gas fuelling SF in the galaxies and the resultant disc fractions, in Fig. \ref{fig:gasStarAlignment}, we show the alignment between the stellar angular momentum (measured within 5~kpc of the galaxy) and gas angular momentum within 5~kpc (top row) and $\RHOST$ (bottom row) as a function of time. The figures show that in the \tng{} simulations, gas within the inner regions of the galaxy is well aligned with the stellar disc over most of cosmic history in all but the \XCXX{} case. In the \vgn{} simulations, the picture is more complex. In the \XXC{}, \XC{} and \XCXX{} cases, we see an early phase where gas is misaligned with the stellar disc (with the gas and stellar angular momentum vectors becoming perpendicular to each other) until $z \sim 0.5-0.7$. At later times however, the trends are similar to those in the \tng{} simulations i.e. the gas is well aligned to the stars. The one exception, the \XCXX{} case, exhibits gas becoming exactly anti-aligned with the stars at $z=0.7$ in the \tng{} simulations leading to the formation of the counter-rotating disc component seen in previous figures. In the \vgn{} simulation, instead of being anti-aligned, the gas angular-momentum is perpendicular to the stars. 

At larger distances, i.e. within $\RHOST$, the gas and stars are not perfectly aligned even at late times. The angle between the angular momentum vectors ranges from $0^{\circ}$ to $60^{\circ}$ for the four simulations with smaller $z=2$ mergers, and for the \XCXX{} simulations, the halo gas counter-rotates at late times. The levels of misalignment are broadly consistent between the two sets of simulations. In contrast, at earlier times ($z>1$) we find significant differences between the \tng{} and \vgn{} simulations with all \vgn{} simulations showing significantly misalignment between stars and gas in the period between $z=1-2$, while the opposite is true for the \tng{} simulations. Since the galaxies, by design, have a quieter merger history at late times, it is expected that the geometry of gas accretion at late times is more dictated by large scale structure where the two simulation codes have better agreement. At earlier times, the gas accretion geometry is instead determined by not only the \targetMerger{} merger, but also a preceding major merger, the orientation of which is strongly dependent on the galaxy formation models.

The results of Figs. \ref{fig:circCompEvolution_newStars}, \ref{fig:circCompEvolution_gas} and \ref{fig:gasStarAlignment} show that the modifications to the \targetMerger{} merger result in similar changes to the gas within the galaxy over its entire history, particularly to the kinematic properties of the gas, thus impacting the kinematics of the resulting galaxies in similar ways. Although the differences in the two galaxy formation models mean that the detailed manner in which the accreted gas is converted into stars can lead to differences in key components of the galaxy, the overall trends appear to be common between the two simulation codes.


\section{Discussion} \label{sec:discussion}

\subsection{How does early merger history impact present-day galaxy size?}

Studies have shown that mergers need not be disruptive events; minor mergers (often defined as those with mass ratios smaller than 1:3) involving gas rich secondary galaxies on gently inspiralling orbits can allow the central galaxies to retain their discs and even continue growing their discs \cite[e.g.][]{Cox2006,Naab2014,Choi2017,Grand2017,Loubser2022,Yoon2022,Graham2023}. Our results also confirm this scenario, with all but the \XCXX{} simulations leading to a disc-dominated galaxy in both sets of simulations, despite the minor merger experienced at \targetMerger{} and an earlier major merger that occurs at $z\sim3$. While the galaxies are temporarily dispersion dominated during the mergers themselves, they are able to quickly re-establish their discs (except in the \XCXX{} case), despite the \targetMerger{} merger being significantly radially oriented in all five scenarios. This suggests that the mass ratio of the merger and the available gas content for future SF is marginally more important in determining the morphology of the resultant merger remnant than the impact parameter.

Increasing/decreasing the mass ratio of the early \targetMerger{} merger results in a smaller/larger MW-like galaxy at $z=0$, while maintaining the mass and size of the encompassing halo. This is directly connected to the galaxies building up a proportionally smaller/larger disc component, allowing the discs to grow bigger. At face value, our results suggest that the \targetMerger{} merger event can have an impact on galaxy sizes nearly 10~Gyr later, which would require a sustained mechanism for the galaxy to either bolster or suppress the formation of a disc component. However, as discussed in \citet{Rey2023,Joshi2024}, when modifying the ICs to alter the \targetMerger{} merger ratio, \textsc{GenetIC} must necessarily alter several other mergers and the overall accretion history of the host halo in order to maintain the halo's $z=0$ mass (a constraint that we imposed). The loss/gain of mass brought in by the secondary system in the targeted merger is compensated for by an increase/decrease in the merger mass ratio of a preceding merger at $z \sim 3$, as well as an extended history of smaller mergers at later times, especially in the \XXC{} and \XVC{} cases. This inter-connectedness of the overall merger history of the galaxies means that we must consider the entire history when determining how the modified ICs lead to larger/smaller galaxy sizes. Additionally, long-range gravitational interactions and correlations implicit in the $\Lambda$CDM power spectrum also cause slight geometrical shifts in the mergers and general accretion. Thus, the sequence of simulations provided by the modifications provides a fascinating controlled testbed for galaxy formation physics, but the results must be interpreted with due consideration for these links between difference aspects of the merger history. 

Figs. \ref{fig:circCompEvolution_allStars}, \ref{fig:circCompEvolution_newStars}, \ref{fig:circCompEvolution_gas} and particularly \ref{fig:gasStarAlignment} together suggest that the modifications made to the ICs result in changes to the gas accretion geometry relative to the disc, with the smaller merger scenarios resulting in more aligned gas accretion and the larger merger scenarios having more misaligned and anti-aligned accretion. This is the ultimate cause of the disc size trends in our suite. The fact that we see similar emergent trends between the two sets of simulations suggests that this connection between merger history and galaxy properties is driven by large-scale environment. The actual sizes are then given an overall normalization by the galaxy formation models. This is in line with the results of \citet{Cadiou2022GasAM} who showed that at $z>2$, gravitational torques are more important than pressure torques, i.e. those from pressure gradients due to internal processes, in setting the AM of the cold gas acquired by galaxies, which in turn determines the stellar AM through SF.

Our results build on previous results finding post-starburst galaxies to be more compact than non-starburst galaxies \citep[such as in][]{Whitaker2012,Yano2016,Almaini2017,Wu2018,Chen2022,Setton2022}. We furthermore show that modifying the merger history of galaxies can in fact alter their sizes by modifying the AM and kinematics of their gas and stars to the present day. While this suggests that the scatter in galaxy sizes at a given mass can be a gauge for the distribution of their merger histories, it also means that such changes in size cannot be attributed to any single merger event.

\subsection{Differences between the \tng{} and \vgn{} simulations} \label{sec:discModelDiffs}

The response of the two galaxy formation models to the modifications to the ICs are similar in terms of the trends in galaxy sizes and kinematics, although the overall normalization of galaxy sizes differs significantly. Even though the trends in properties are similar, detailed examination reveals that the two codes arrive at these trends along somewhat different pathways. Here we discuss some of the differences between the two codes that contribute to the distinct behaviours of the corresponding galaxies.

\subsubsection{Bulge fraction and AGN feedback}
As Fig. \ref{fig:circDistributionZ0} shows, the \vgn{} galaxies all have a higher bulge proportion than the \tng{} galaxies. This can also be seen in the fiducial surface brightness profiles in Fig. \ref{sec:fiducialComp}, where the VG galaxy has a significantly brighter central region within $\sim 200$~pc, while the surface brightness in the intermediate regions of $0.5-5$~kpc are comparable to each other. The \vgn{} galaxies' high concentration is also reflected in a higher kinematic bulge fraction, which makes their sizes systematically smaller than the \tng{} galaxies. These differences hold even at higher redshift, as seen in Fig. \ref{fig:circCompEvolution_allStars}. As mentioned earlier, a significant factor in determining galaxy sizes is the presence of AGN feedback in the TNG model, which is not included in the VG model. The AGN feedback (especially in kinetic mode) results in the ejection of low-angular momentum, dispersion-dominated gas from the central regions of the galaxy, thereby suppressing bulge growth \citep[e.g. ][]{Grand2017}. This is one of the contributing factors to the \vgn{} galaxies being significantly smaller than the \tng{} galaxies. However, as seen in Fig. \ref{fig:sizeMassRelation}, both sets of galaxies are in agreement with observational results from L16 for the size-mass plane, when compared with galaxies of matched morphologies. Our results also suggest AGN feedback may be an important contributor to the scatter in the size-mass relation, both in simulations and observations.

\subsubsection{Gas accretion geometry}
While the absolute sizes are strongly affected by the different feedback processes in the \vgn{} and \tng{} models, the trends remain similar because they are predominantly driven by gas accretion geometry (see Section \ref{sec:gasStarAlignment}). In particular, we show how similar levels of (mis)alignment between the gas and stellar angular momentum at late times in both sets of simulations can explain the similar trends in galaxy sizes we have found. The misalignment of gas and stars can be caused by two factors: i) changes in large scale gas inflows and ii) rotation of the stellar disc due to external torques. To determine the importance of both these factors, we have considered the rotation of the stellar and gas angular momentum vectors relative to their own directions at $z=0$. For brevity we do not show the relevant plots here, but we find that in the \tng{} simulations, gas within $\RHOST$ shows little rotation of its angular momentum over the entire history of the galaxies. (The only exception is the \XCXX{} galaxy at early times, where the gas angular momentum vector rotates by nearly 180$^{\circ}$). As well as this stability in the gas accretion, the stellar disc angular momentum rotates only moderately, remaining largely in the same direction as at $z=0$. Together, this allows the gas and stellar components to be well aligned in the \tng{} simulations over most of cosmic time, which in turn promotes disc growth.

In the \vgn{} simulations on the other hand, there is significant rotation in both the gas and stellar angular momentum vectors; importantly such rotations do not occur synchronously in the fiducial and two extreme merger scenarios, resulting in the misalignments seen in Fig. \ref{fig:gasStarAlignment} at early times. This is not the case for the two intermediate galaxies, \XVC{} and \XCX{}, which show relatively less stellar rotation that is mostly synchronous with the gas rotation, consistent with the alignment seen in Fig. \ref{fig:gasStarAlignment}. It is possible that, in addition to the misalignment of incoming gas and the existing stellar disc, the torquing and rotation of the disc itself may promote bulge growth by disrupting the orbits of disc stellar particles into random orientations.

Thus, although the trends in the star-gas alignment are similar between the two codes, they are achieved along distinct pathways, which also drive the distinct kinematic make-up of the galaxies discussed above. If higher amounts of torquing and stellar rotation of the stellar disc leave signatures in the $z=0$ galaxies as warps or streams, this may be an avenue in distinguishing between such evolutionary pathways. Understanding why the two simulation codes produce these different pathways will be a topic of future investigation.

\subsubsection{ISM modelling and SF}
Another crucial factor is the treatment of the ISM in both models. The VG model allows gas to cool down to 10~K, thus resolving and forming stars in significantly denser regions than the TNG model. Additionally, the different SF thresholds in the two models imply that SF in the TNG model can occur at much larger radii than in the VG model, which would also explain the larger TNG discs. We also see the impact of the resolved ISM in Figs. \ref{fig:circCompEvolution_newStars} and \ref{fig:circCompEvolution_gas}. The kinematic distribution of the in-situ SF largely follows that of the gas and is similar between the two codes. The SF is dominated by gas that is rotationally-supported, but in the \tng{} case, there is also significant SF in the bulge component. In the \vgn{} case, by contrast, we see that over the period between $z=1-2$, stars do not form directly from the bulge, even though there is a significant component of non-rotationally-supported gas. In the \vgn{} \XCXX{} case, we only see appreciable SF in the `bulge' component in the last two Gyr, when the gas in this component accounts for the majority of gas in the galaxy. These results must be reconciled with the fact that the \vgn{} galaxies have higher bulge fractions than the \tng{} galaxies. The apparent tension demonstrates that another process contributes significantly to the building of the bulge in \vgn{}, i.e. the heating up of disc stars into dispersion-dominated orbits. We will explore this in detail in a future paper.

\subsection{The role of stochasticity}
When interpreting the trends found in our results, the stochasticity in the two models is an important factor to consider. Figs. \ref{fig:sizesToday} and \ref{fig:sizeEvolution} showed that there is stochastic scatter in the TNG galaxy sizes, likely stemming from the SF and AGN feedback prescriptions, as well as the computational infrastructure (e.g. see \citealt{Genel2019} for the IllustrisTNG model itself, and \citealt{Keller2019,Genel2019,Davies2021,Davies2022,Borrow2023} for other galaxy formation models). In Figs. \ref{fig:circCompEvolution_allStars}, \ref{fig:circCompEvolution_newStars} and \ref{fig:circCompEvolution_gas} however, this scatter does not have a large impact on the relative proportion of `disc' and `bulge' stellar and gas components or SF, indicating that the modelling of internal physical processes driving galaxy evolution is fairly robust against stochasticity in the TNG model. We have also confirmed that the halo mass growth is nearly identical and shows no stochastic uncertainty after $z=4$. Instead, we find that the uncertainty in galaxy sizes is due to scatter in the gas accretion geometry, and also in the growth of the BH. The galaxy that grows the largest in size among the three runs of the fiducial ICs has the lowest stellar mass, shows greater alignment between the stars and gas out to $\RHOST$ and hosts the least massive BH. This further strengthens the correlation between smaller BH masses and larger disc sizes in the \tng{} simulations, as discussed in Section \ref{sec:agn}. It also explains the behaviour of the intermediate simulations SM1 and LM1, which do not precisely follow trends defined by the two extreme merger cases. In particular, their gas accretion geometry and BH mass are seen to behave in an opposite manner to the \XXC{} and \XCXX{} simulations i.e. the \XVC{} BH is more massive and the \XCX{} BH is less massive than the fiducial galaxy. The \XXC{} and \XCXX{} scenarios are sufficiently different from the \XC{} scenario that the stochastic scatter is subdominant to the trends.

While we have not quantified the stochasticity in the VG model, it is likely that a similar degree of uncertainty applies to the \vgn{} results. Results from the EDGE simulations \citep{Rey2019,Agertz2020}, which use a similar galaxy formation model to VINTERGATAN, have shown that there can be noticeable stochastic scatter in galaxy properties, but that it is usually sub-dominant to the modifications made by \textsc{GenetIC} \citep{Pontzen2021}. Previous work on other galaxy formation models have also shown that stochasticity plays a non-negligible role in driving scatter in galaxy evolutionary histories and resultant properties (e.g. \citealt{Davies2021,Davies2022,Borrow2023} for the EAGLE model of \citealt{Crain2015,Schaye2015} and \citealt{Keller2019} for the GASOLINE2 model of \citealt{Wadsley2004,Wadsley2017}). Thus, the stochastic uncertainty generated within the galaxy formation simulations sets a limit to what physical interpretations can be made from them, whether within a volume simulation or a more controlled study such as ours. Nonetheless, this limit appears to be sub-dominant to the overarching causal trends of interest in most works to date, including our own.


\section{Summary \& conclusions} \label{sec:conclusions}

In this paper, we explore the long-term impacts of modifying the early merger history of a MW-mass galaxy on its present-day properties, particularly to understand how it can alter its kinematic make-up and therefore its size. We present the PARADIGM project, consisting of the VINTERGATAN-GM suite of zoom-in cosmological hydrodynamical simulations of MW-mass haloes \citep{Rey2022,Rey2023} produced with the VINTERGATAN galaxy formation model (referred to as \vgn{}), and accompanied by a new complementary suite of simulations generated with the IllustrisTNG model, named IllustrisTNG-GM (referred to as \tng{}). The two sets of simulations evolve the same set of five ICs, which were generated with the \textsc{GenetIC} algorithm, consisting of a fiducial version and four `genetically modified' versions that alter the mass ratio of a \targetMerger{} merger, while maintaining the halo mass at $z=0$. The GM ICs allow us to isolate the impact of varying merger histories on resultant galaxy properties while controlling for other factors, as much as possible, that may also affect the galaxy. 

However, it is not logically possible to alter only a single merger without affecting other parts of the merger history to compensate for the modifications, and so our analysis has carefully considered the range of factors that changes across the controlled suite. Rather than focusing solely on one-to-one comparisons between the corresponding galaxies -- which may differ significantly due to the different physical ingredients implemented in each code as well as the differing resolutions achieved in the simulations -- the GM approach allows us to compare the trends in galaxy properties as a function of \targetMerger{} merger mass ratio. With this approach, we examine how $z=0$ galaxy sizes, i.e. half-light and half-mass radii, depend on early merger history, and how the two galaxy models respond similarly to the modified ICs despite significant differences in their numerical and physical implementations. Our main results can be summarised as follows:

\begin{itemize}
    \item We find a clear trend of increasing $z=0$ galaxy size with decreasing \targetMerger{} merger mass ratio due to accompanying changes in the overall accretion history. The trends are stronger when considering half-mass radius, but are nonetheless also significant when considering the half-light radius. The size differences are larger than the differences in stellar mass, so that the GMs contribute to the scatter in the galaxy size-mass relation rather than displace the galaxies along the relation. Both sets of galaxies are found to be consistent with observations, albeit with the \tng{} galaxies matching the distribution of discier Sab-Scd galaxies (and being mildly larger than average) and the \vgn{} galaxies matching the distribution of S0-Sa galaxies (and being marginally smaller than average). These trends are significant even when accounting for stochastic uncertainty in the TNG model (stochasticity in the \vgn{} model has not been quantified due to computational cost).
    \item The trends in half-light radius seen at $z=0$ are established by $z\sim1$ in the \tng{} simulations (approximately 2-3 Gyr after the end of the \targetMerger{} merger) and at $z\sim0.5$ in the \vgn{} simulations, highlighting the importance of the entire merger history of the galaxies in establishing the $z=0$ trends.
    \item The evolution of the galaxies in the size-mass plane confirms that the trends in galaxy size are seen even at fixed stellar mass. The evolution in this plane is characterised by a mild contraction phase, where galaxy size decreases with increasing stellar mass, followed by an expansion phase, where size increases with increasing stellar mass (though much more significantly for the \tng{} galaxies than the \vgn{} galaxies). The transition between the phases occurs at $z \sim 1.5$ in the \tng{} simulations (soon after the end of the \targetMerger{} merger) and $z \sim 0.5$ in the \vgn{} simulations. The galaxy sizes are much more in agreement between the two sets of simulations during the contraction phase.
    \item The kinematic make-up of the stellar particles in the galaxies at $z=0$ is characterised by a dominant disc component and secondary but important bulge component, in all but the \XCXX{} case. In \XCXX{}, the bulge component is at least of equal mass, if not more massive than the disc component. The \tng{} galaxies are characterised by a much more massive, kinematically colder disc component, while the \vgn{} galaxies host a significant bulge component. Nonetheless, we see similar trends in both sets of simulations with disc fraction increasing with decreasing \targetMerger{} merger mass ratio.
    \item By following the evolution of the disc, bulge and counter-rotating disc components, we confirm that the growth of a larger/smaller galaxy is linked to the build up of a larger/smaller disc versus bulge component. In the \tng{} \XCXX{} case, a counter-rotating disc component accounts for nearly 10 per cent of the stellar mass and its presence is linked to the galaxy being bulge-dominated at $z=0$.
    \item In both codes, the kinematic make-up of in-situ SF matches that of the gas within the galaxy. The changes in the total disc, bulge and counter-rotating masses are driven primarily by in-situ SF, but there is significant contribution to the build-up of the bulge component from disc stars whose orbits become more disordered over time.
    \item In-situ SF predominantly occurs within the disc component, except in the \tng{} \XCXX{} case, where SF is dominated by the counter-rotating disc component after $z \sim 0.7$. In the \vgn{} \XCXX{} case, we also see SF in the bulge and counter-rotating disc component, albeit at a lower level than in TNG. 
    \item The similarity in the response of the two codes to the modified ICs can be attributed to comparable gas accretion geometries, with the gas and stars in the inner region being well aligned, especially at late times, with the exception of the \XCXX{} galaxy, which becomes bulge-dominated. These results suggest that the two galaxy formation models transport gas and its associated angular momentum from the halo (and beyond) to the galaxy in similar ways, despite a comparison of single galaxies between the codes showing major differences.
\newline

Our new approach to comparing galaxy formation models can provide insights into the connection between the input physics ingredients and the emergent galaxy evolution and observable properties. The TNG and VG models represent contrasting and complementary approaches to simulating galaxy formation. The TNG model takes a more top-down approach, incorporating effective prescriptions for as many physical mechanisms as computational costs will allow, and choosing most free parameters to reproduce key $z=0$ scaling relations. This approach is powerful in understanding the population of galaxies in the Universe, what processes are key to their evolution and what the rates of their occurrence are. On the other hand, the VG model takes a more bottom-up approach, incorporating a smaller number of key physical ingredients as much as possible from first-principles, thus facilitating the detailed understanding of these processes and their implications for galaxy formation. The two approaches therefore provide complementary insights on galaxy formation and evolution. In the present work, we have shown how they converge on several emergent phenomena, despite producing apparently very different galaxies. 

More broadly, our comparison highlights the complex interplay between large-scale gravitational collapse and internal evolutionary and feedback processes. The existing literature highlights significant uncertainty in our picture of galaxy formation physics either at a population level or at the level of individual galaxies. The new type of comparison presented here, in which we compare and contrast the response of different galaxy formation physics implementations to systematic changes in a galaxy's formation history, provides a complementary third perspective. In future work, we will explore several other aspects of the galaxies in our simulation suite e.g. SFHs, metallicity gradients, the intrinsic alignments of stars and gas and the connection to large-scale structure.  
    
\end{itemize}


\section*{Author contributions}

\textbf{GJ}: Conceptualization, data curation, formal analysis, investigation, writing -- original draft (lead). \textbf{APo}: Conceptualization, funding acquisition, methodology, resources, writing -- original draft (support). \textbf{OA}: Conceptualization, data curation, methodology, resources, writing -- review \& editing. \textbf{MR}: Conceptualization, data curation, investigation, writing -- review \& editing. \textbf{JR}: Conceptualization, writing -- review \& editing. \textbf{APi}: Conceptualization, methodology, writing -- review \& editing.


\section*{Acknowledgements}

We thank Florent Renaud for valuable discussions and insights while writing this paper. We thank Volker Springel for his insights in the conception of this project and the IllustrisTNG team for developing the model and sharing the code used in this work. We also thank the anonymous referee for their insightful comments that were useful in improving the initial manuscript. This project has received funding from the European Union’s Horizon 2020 research and innovation programme under grant agreement No. 818085 GMGalaxies. This study used computing equipment funded by the Research Capital Investment Fund (RCIF) provided by UKRI, and partially funded by the UCL Cosmoparticle Initiative. OA acknowledges support from the Knut and Alice Wallenberg Foundation, the Swedish Research Council (grant 2019-04659), and the Swedish National Space Agency (SNSA Dnr 2023-00164). MR is supported by the Beecroft Fellowship funded by Adrian Beecroft. We acknowledge PRACE for awarding us access to Joliot-Curie at GENCI/CEA, France to perform the simulations presented in this work. Computations presented in this work were in part performed on resources provided by the Swedish National Infrastructure for Computing (SNIC) at the Tetralith supercomputer, part of the National Supercomputer Centre, Linköping University. JR would like to thank the STFC for support from grants ST/Y002865/1 and ST/Y002857/1.


\section*{Data Availability}

The data underlying this article will be shared upon reasonable request to the corresponding author.



\bibliographystyle{mnras}
\bibliography{VGM_vs_TNG_sizes_and_morphology} 




\appendix

\section{$\RHALF$ as a measure of galaxy size} \label{sec:appendixSizes}

\begin{figure*}
    \includegraphics[width=\linewidth]{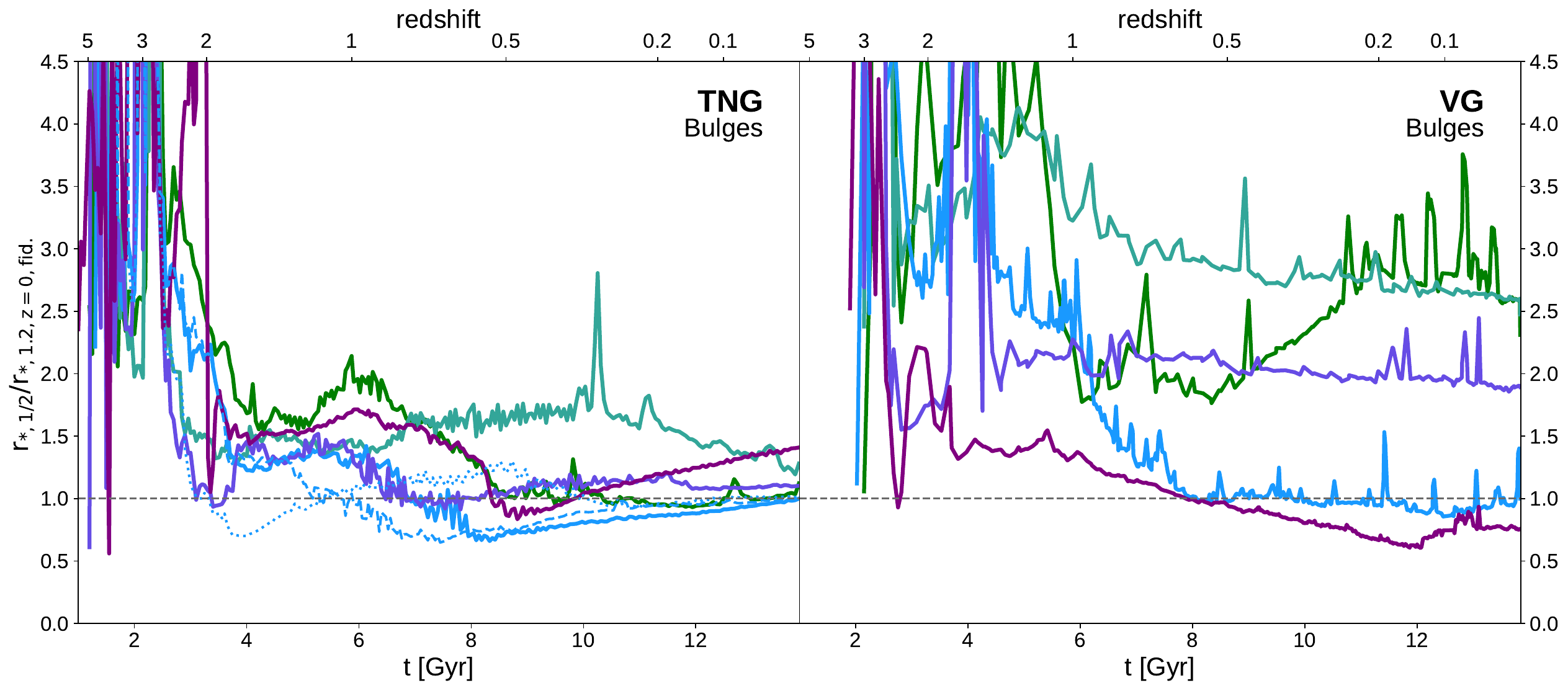}
    \includegraphics[width=\linewidth]{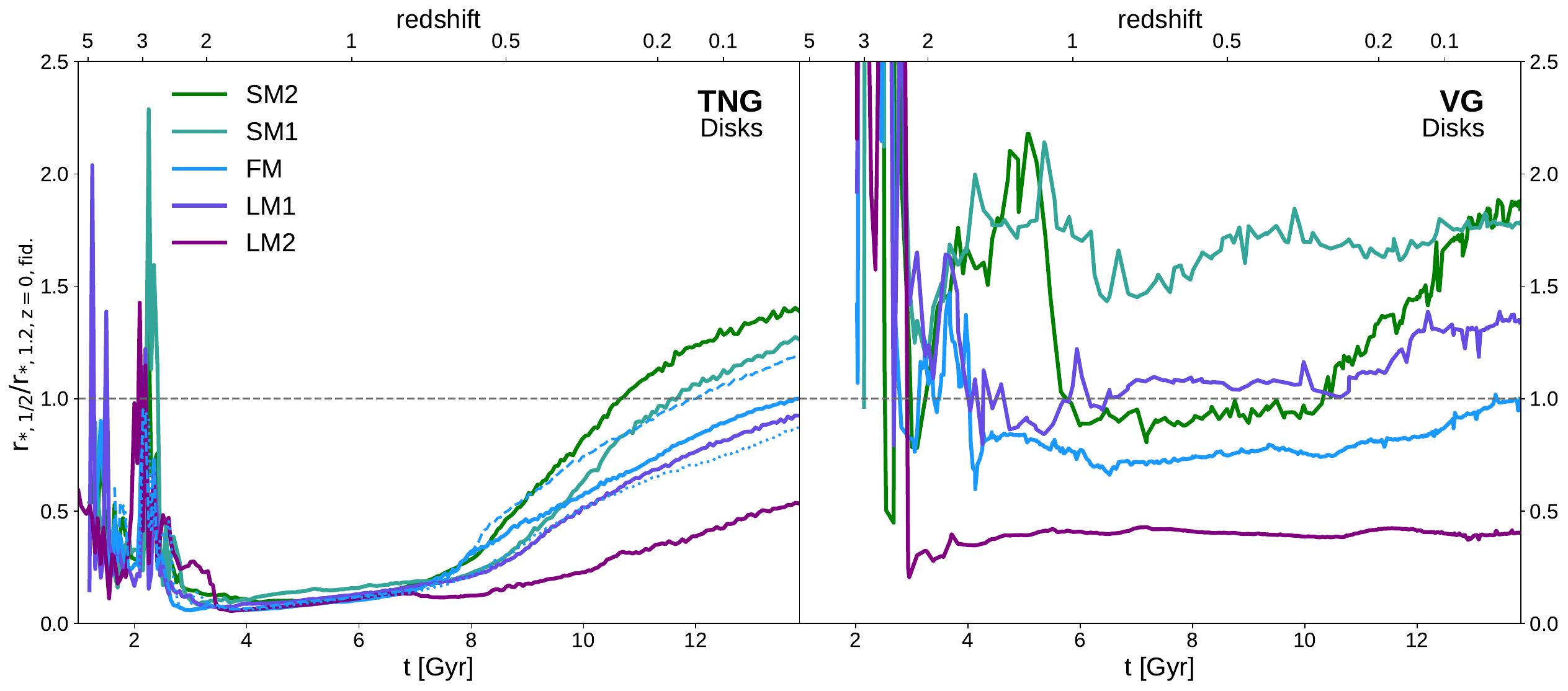}
    \caption{Evolution of the half-mass radius of the bulge (top) and disc (bottom) components separately, as in Fig. \ref{fig:sizeEvolution} for the overall sizes of the galaxies. Bulge/disc mass profiles were constructed by fitting the circularity parameter distributions in log radial bins of width 0.1 dex and then used to calculate the corresponding half-mass radius at each timestep. The sizes here have been normalized by the $z=0$ fiducial bulge and disc half-mass radius. For the \tng{} galaxies, these values are 0.55 and 6.28 kpc respectively, and 0.56 and 0.88 for the \vgn{} galaxies. The evolution of the \tng{} sizes (i.e. $\RHALF$) seen in Fig. \ref{fig:sizeEvolution} largely reflects the disc sizes of the galaxies, which are significantly larger than their bulges after $z\sim0.7$ (but smaller or comparable before this). For the \vgn{} galaxies, the disc and bulge sizes are comparable at all times, with the bulges even being larger than the disc in some cases. Therefore, their size evolution cannot be attributed to a single component, but rather reflects the growth of the galaxy as a whole.} \label{fig:sizeEvolutiondiscBulge}
\end{figure*}

It is clear that the \tng{} and \vgn{} models produce morphologically distinct galaxies, with the \tng{} galaxies hosting kinematically cold discs and relatively smaller bulges, whereas the \vgn{} galaxies show significant overlap in the circularity parameter distribution of the `disc' and `bulge' components. In the \tng{} galaxies, $\RHALF$ largely measures the size of the discs themselves, whereas in the \vgn{} galaxies, it likely pulled to lower values due to their prominent bulges. To understand how this parametrization of galaxy size affects our results, we have considered the results of Fig. \ref{fig:sizeEvolution} separately for the disc and bulge components. To do so, we employ the strategy described in Section \ref{sec:kinDecompZ0} to measure disc/bulge masses in log radial bins of 0.1 dex, construct the corresponding mass profiles and then measure a half-mass radius for each separately. The results are shown in Fig. \ref{fig:sizeEvolutiondiscBulge}, as in Fig. \ref{fig:sizeEvolution} for the overall galaxy sizes. Note that here, we normalise by the $z=0$ fiducial bulge and disc sizes. 

For the \tng{} simulations, we confirm that the trends in Fig. \ref{fig:sizeEvolution} are indeed dominated by the size of the discs, and are in fact stronger when considering disc sizes alone. As with the overall sizes, the disc sizes are largely consistent with each other and show moderate growth until $z \sim 0.7$, after which we see accelerated growth in the discs. It should also be noted that until $z \sim 0.7$, the \tng{} discs are in fact smaller than or comparable to the \vgn{} discs, and only become larger due to this rapid growth at later epochs. At $z=0$, the disc half-mass radii range from $3.5-8.5$~kpc and are nearly always larger than the overall half-mass radii. On the other hand, there is no discernible correlation between merger history and bulge sizes, which range between $0.5-0.8$~kpc at $z=0$. The bulges experience a period of moderate contraction between $z=0.5-1$, after which time the bulge sizes are either constant or show moderate growth to present day.

The case of the \vgn{} simulations is markedly different. Firstly, we find that the disc and bulge sizes are comparable to each other, with their $z=0$ values ranging from $0.3-1.6$~kpc. While the disc sizes largely remain constant or show mild growth after $z=1$, the bulges show mild contraction over this entire period. The one exception is the \XXC{} simulation, which shows significant and steady growth in both disc and bulge sizes, which are of comparable size throughout its history. 

The different behaviour of the two codes is partly due to the markedly different distributions of $\epsilon$ which we use to separate the bulge and disc components. The \tng{} galaxies have a proportionally more dominant disc component (except in the \XCXX{} case) than the \vgn{} galaxies. Therefore, the $\RHALF$ measurements for the latter have a larger contribution by the bulge component. However, these results confirm that although $\RHALF$ does not characterise the \tng{} and \vgn{} discs in the same way, there are no qualitative differences between its evolution and that of disc sizes alone.


\bsp	
\label{lastpage}
\end{document}